\documentclass[a4paper, 12pt]{article}
\usepackage{lipsum}
\usepackage{multicol}
\usepackage{float}
\usepackage{amsmath}
\usepackage{amsfonts}
\usepackage{subcaption}
\usepackage{amssymb}
\usepackage{graphicx} 
\usepackage{hyperref}
\usepackage{geometry} 
\usepackage{setspace}
\usepackage{authblk}
\usepackage{cite}
\usepackage{fix-cm}
\usepackage{url}
\usepackage[utf8]{inputenc}
\usepackage{tabularx}
\usepackage{multirow}
\usepackage{caption}
\usepackage{array}
\usepackage{mathtools}
\usepackage{amssymb}
\usepackage{textcomp} 
\usepackage{geometry}

\date{}

\begin{document}

	\title {Investigating the effect of adaptive optimal control function in epidemic dynamics: predictions and strategy evolution based on SIR/V game theoretic framework}

	\author[1]{\small Nuruzzaman Rahat\thanks{Email: nuruzzamanrahat1738@gmail.com }}
	
          \author[1]{\small Abid Hossain\thanks{Email: abid28266@gmail.com}}
	
	\author[1]{\small Muntasir Alam\thanks{Corresponding author: Muntasir Alam, muntasir.appmath@du.ac.bd}}

	\affil[1]{\footnotesize Department of Applied Mathematics, University of Dhaka, Dhaka-1000, Bangladesh}
	

	\maketitle
	
	\begin{abstract}
	\noindent In this paper, we consider an adaptive optimal control problem for an SIR/V epidemic model with human behavioral effects.We develop a model where effective management of infectious diseases are monitored by the means of non pharmaceutical interventions.This study develops an adaptive optimal control function within an SIR/V framework embedding a non cooperative game theoretic mechanism to capture the dynamic interplay between individual vaccination behavior and population level transmission. We derive analytical expression for the optimal control trajectory under resource constrain and heterogeneous susceptibility and we validate our model using numerical simulations,calibrated with the real world epidemic parameters. We find that for the adaptive optimal policy for a generally known SIR/V model depending on the game theoretic epidemic state leads to substantial reduction in expenses compared to non adaptive policies. Moreover, our results demonstrate that, adaptive strategies significantly outperform the static policies by achieving lower peak infections and faster epidemic extinctions while evolutionary game dynamics identify critical behavioral thresholds that drive strategy evolution and inform timely policy adaptation

		 \par

	\end{abstract}

	\section*{Keywords}
	Epidemic modeling, Evolutionary process, Strategy Updating, Vaccination cost, Adaptive optimal control, Social payoff;
	
	\section*{Highlights}

	\begin{itemize}

		\renewcommand{\labelitemi}{$\checkmark$}

		\item Develop and compare adaptive optimal control strategies for early outbreak mitigation in an SIR/V framework, integrated with evolutionary game-theoretic strategy updating rule.
		
		\item Analyzes trade-offs between intervention intensity and socio-economic burden.
		
		\item This study shows adaptive strategies reduce infections and costs more effectively.
		
		\item Illustrates how non-pharmaceutical interventions complement vaccination strategies.
		
		\item The proposed model reveals that adaptive control maintains system stability under varying epidemic conditions.
		
		\item This research emphasizes the interplay between individual choices, societal behavior , and policy efficiency. 
		
	\end{itemize}

	\section{Introduction}\label{introduction}
	\noindent Mathematical modeling has allowed for the understanding and control of infectious disease over many decades, presenting a way to analyze transmission dynamics and the effects of different interventions. Many investigations have explored concepts such as vaccination, social distancing, quarantine, and treatment, often with optimal control theory in order to find the most effective and cost-efficient interventions. The recent literature, however, has several limitations. Many studies examine a single intervention alone, ignoring the complex interplay among different public health interventions. Moreover, the role of human behavior, e.g., a community's awareness or reaction to public health messaging, is represented simplistically or ignored altogether. While there has been some research taking into account the potential for using game theory to model strategic decision-making during epidemics, no research has explored the application of evolutionary game theory together with the adaptive control approach for adaptively controlling intervention policies to co-evolve with population behavior. Finally, while computational tools for the solution of optimal control problems have matured, their application to multi-variable complex epidemiological systems with realistic real-world constraints remains a challenging task. The above limitations are overcome by this research through the creation of a new optimal control framework. We construct an extensive model with multiple interventions, i.e., vaccination with adaptive social awareness campaigns, to reflect a more realistic control environment. Our new approach makes explicit the representation of the role of public health messaging in influencing individual behavior and, in turn, disease transmission. In addition, we apply state-of-the-art numerical methods to solve this difficult, multi-control optimal control problem efficiently and provide a valuable and robust tool to inform integrated public health policy. 
	
	\noindent The application of optimal control theory to biological and epidemiological models has been an area of significant research. Optimal control problems, which aim to find control actions maximizing or minimizing the performance of a system over time, are generally challenging to solve. A number of approaches have been developed to solve such problems. A multiple shooting algorithm, for instance, has been proposed to solve optimal control problems directly, offering a numerical approach for calculating optimal trajectories and controls \cite{Bock1984}. As well, limited-memory algorithms have been introduced for bound-constrained optimization, offering efficient solutions to large-scale problems in such cases \cite{Byrd1995}. The challenge in modeling and forecasting the spread of infectious diseases, as has been the case in the COVID-19 pandemic scenario, highlights the application of robust mathematical tools \cite{Bertozzi2020}. Solution approaches such as the forward-backward sweep approach are used to solve such optimal control challenges, and convergence characteristics have been studied to offer robust solutions \cite{McAsey2012}. For complex systems, such as in systems biology, the application and speed-up of optimal control methodologies are vital to practical realization \cite{Sharp2021}.
	
	\noindent Mathematical modeling of infectious diseases has existed for decades, with previous research serving as the basis for present-day epidemiological research \cite{Hethcote2000}. Optimal control has in this regard been applied to a wide range of biological models, such as chemotherapy for HIV, where optimal control theory was employed to determine the best treatment strategies \cite{Kirschner1997}. Optimal control has also been researched towards curtailing the spread of a novel virus outbreak, with the research illuminating intervention strategies \cite{Smirnova2024}. Construction of the next-generation matrix is a crucial building block in compartmental epidemic models, which enables investigation into disease transmission dynamics \cite{Diekmann2010}. IPM principles have also been applied in a forest ecosystem, showing the application of management strategies in managing biological systems \cite{Sweeney2023}.
	
	\noindent Vaccination is one of the primary interventions that have been utilized in the control of epidemics. Research has approximated the percentage of vaccination coverage thresholds to achieve herd immunity for diseases like SARS-CoV-2 \cite{PlansRubio2022}, and estimation of herd immunity thresholds from current epidemics \cite{Aguas2020}. The performance of different vaccination strategies, e.g., imperfect vaccination versus protection against contagion, have also been studied based on statistical mechanics models \cite{Kuga2018}. Resource allocation and distribution, e.g., vaccines, during a pandemic is an optimal control problem. A review of the quantitative models for vaccine distribution and allocation, including equity, hesitancy, and the COVID-19 pandemic, indicates the need for enhanced models to inform public health policy \cite{Blasioli2023}.
	
	\noindent Aside from medical measures, adaptive human behavior may significantly influence the course of an epidemic, and the integration of such behavior into epidemiological models is required to provide realistic predictions and inform policy in due time \cite{Fenichel2011}. Social distancing, for example, can be analyzed by using optimal control and differential game theory to provide efficient strategies to curb infectious disease propagation on networks \cite{Dashtbali2020}. A review of game-theoretic models in the area provides a broad overview of the influence of human decision-making on epidemic spreading \cite{tanimoto2021sociophysics, tanimoto2021social, Huang2022, Alam2019three, alam2020based,alam2021abrupt,sushmit2024dynamic, kulsum2024modeling, akter2024silico}. A game-theoretic model has also been used to model social distancing as a response to an epidemic with a focus on strategic interaction between individuals \cite{Reluga2010}.
	
	\noindent Various optimal control techniques have been developed for various epidemic models. For instance, optimal control of a finite quarantine SIR epidemic model has been studied in order to obtain the optimal isolation policies \cite{Balderrama2022}. Differential game theory in combination with the SEIR model has also been used in order to obtain optimal infectious disease control policies \cite{Alabdala2024}. In real-world scenarios, uncertainty in parameters like transmission and treatment rates may affect the outcomes. Optimal control of the SIR model with such uncertainty included has been studied in order to obtain more robust control policies \cite{Gatto2021}. The computational task of solving such problems typically involves optimization of complex functions, which can be enhanced by new methods of approximating Hessian matrices, like through Bayesian inference for quasi-Newton optimization for stochastic optimization \cite{Carlon2022} or through exploration of the analytical structure and sparsity of the hypervolume indicator Hessian matrix \cite{Deutz2023}. The principles of optimal control theory are well established, and they constitute the theoretical foundation for such applications \cite{Kirk2004}. To speak in detail about the application of optimal control to biological systems, a specialty textbook gives a thorough treatment of the topic \cite{Lenhart2007}.
	
	\noindent Mathematical modeling of infections has a long history, and seminal works laid the groundwork for modern epidemiological research. Kermack and McKendrick formulated the classical SIR model, the backbone of compartmental epidemic models \cite{Kermack1927}. This has been accompanied by classical mathematical texts on control and dynamics of infections \cite{Anderson1991} and population biology \cite{Brauer2012}. The basic reproduction number, $R_0$, is one of the most significant epidemiological quantities, and calculation and properties of $R_0$ have been addressed in numerous studies \cite{Driessche2008}. Mathematical models are a useful tool to analyze disease transmission dynamics, such as sexually transmitted diseases \cite{Boily1997}.
	
	\noindent The application of optimal control theory to biological and epidemiological models has been a significant area of study. Optimal control problems, optimizing or minimizing system behavior over time using control measures, are often challenging to solve as outlined in the seminal work of Pontryagin \cite{Pontryagin2018}. An introductory text to the subject is the work of Neilan and Lenhart \cite{Neilan2010}. Various numerical and computational methods have been developed for the problems, with early computing methods laying down the foundation for advanced methods \cite{Balakrishnan1968}.
	
	\noindent Optimal control has been applied to a wide range of biological models and interventions for disease control. Examples include optimal control of malaria by vector interventions and drug treatment \cite{Khamis2018, Wenger2013}, and tuberculosis control by cost-effectiveness analysis \cite{Rodrigues2014}. The literature has further considered control of vector-borne diseases using treatment and prevention \cite{Blayneh2009} and optimal control approaches for Ebola using quarantine and vaccination \cite{Ahmad2016}. The COVID-19 pandemic caused yet another wave of modeling activity. Models specific to this environment were created, including a two-region SEIR model for the island of Ireland \cite{Grannell2020}, a delayed SEIR model \cite{Liu2024}, and models for specific regions like Nigeria with optimal control interventions \cite{Abioye2021}. Optimal control interventions for pandemic influenza control have also been formulated in the same way \cite{Kim2017}. Further research on the optimal control of fractional-order models \cite{Mahata2022} and limited resource models \cite{Hansen2011} is pushing the field further.
	
	\noindent The intersection of game theory and network modeling gives precise accounts of decision-making in decentralized environments within the context of epidemics. Individual-level vaccination strategies, for example, can be analyzed through a game theory formulation on a network, which captures the effects of individual-level choices to derive public health repercussions \cite{Reluga2013}. More abstractly, linkages between differential games and optimal control have been the subject of ongoing investigation, with monographs capturing generalized dynamics of first-order partial differential equations with applications in such fields \cite{Melikyan2012}. Numerically, solutions to optimal control problems in practice often rely on advanced numerical methods, such as nonlinear programming, which must be put into practice \cite{Betts2010}.
	
	\noindent Beyond medical interventions, adaptive human behavior and public health campaigns are critical control measures. Studies have modeled awareness programs by media as an optimal control strategy to influence disease dynamics \cite{Misra2015}. Social distancing, for example, has been analyzed using optimal control to find effective solutions for mitigating the spread of infectious diseases \cite{Saha2023}. The role of human mobility in disease transmission is a significant factor in modeling and has been extensively reviewed \cite{Lessani2024}. The field of vaccination modeling has seen a focus on strategies like pulse vaccination within the SIR framework \cite{Shulgin1998} and the optimization of campaigns considering societal characteristics \cite{Lee2025}.

	\section{Methodology}
	\label{Methodology}
	
	We take the SIR model as the foundational framework and extend it to account for more complex dynamics, such as SIR/V or SVIR systems and with the implementation of the optimal control strategy for infectious population mitigation. In doing so, we are aiming to capture the real-world scenario where we are considering the control strategy being implemented while vaccination procedures are being employed. This extension is crucial for analyzing the impact of controlling the population before beginning the vaccination campaigns in diverse population settings \cite{Kirschner1997,Carlon2022}. Additionally, we consider an infinite, well-mixed population to simplify our assumptions and focus on the dynamics of disease spread and effective control.
	
	\subsection{Effectiveness model }
	
	In the controlled vaccinated population, individuals can be divided into two categories: those who acquire perfect immunity after vaccination and those who do not gain immunity, either due to vaccine failure or insufficient immune response. The effectiveness of the vaccine, denoted by $e$ $(0 \leq e \leq 1)$ , represents the probability that a vaccinated person gains immunity, while the vaccination coverage $x$ represents the proportion of the population that is vaccinated. Therefore, the fraction of vaccinated individuals who successfully gain immunity is given by $ex$, while the remaining fraction of non-immune individuals in the vaccinated population is $(1 - ex)$.\\
	
	\noindent At equilibrium, when the epidemic has run its course ($t = \infty$), we can express the final epidemic size (FES) R as a function of both $x$ (the vaccination coverage) and the time $t$. This allows us to estimate how many individuals in the population remain uninfected due to either natural recovery or successful immunization, which is vital for predicting long-term herd immunity and understanding the thresholds needed to prevent future outbreaks.\\
	
	$R(x, \infty) = (1 - ex)\left(1 - \exp\left[-R_0 R(x, \infty)\right]\right)$\\
	
	\noindent The primary goal of our study is to explore control strategies that can be implemented during the early ascending phase of an outbreak, before more comprehensive measures such as vaccines and antiviral medications become available. In the classic SIR/V (Susceptible–Vaccinated– Infectious–Recovered) model, these early interventions are crucial for managing the spread of the virus. Common mitigation strategies widely employed during the COVID-19 pandemic include physical distancing, enhanced personal hygiene, mask-wearing, and public awareness campaigns. These measures aim to ``flatten the curve'' by reducing the rate of new infections, thereby lowering the overall burden on healthcare systems and decreasing virus-related mortality.\\
	
	\noindent In the context of the SIR/V model, control strategies such as vaccination (even if limited early on) and other non-pharmaceutical interventions can be represented by a time-dependent control function $u(t)$. This control modifies the dynamics of the model by influencing the transmission rate, contributing to the reduction of new cases during the outbreak's critical early phase.By incorporating these controls, we can assess how effectively they mitigate the spread and their potential impact on the overall course of the epidemic, which typically takes the usual form of	$\frac{dx}{dt} = f(x,u)$ \\
	
	The SIR/V model [3][13] we implemented to describe such a scenario is
	\[
	\begin{aligned}
		f_1(x,u) &= -\beta(1 - u(t)) S(x,t) I(x,t)  \\
		f_2(x,u) &= -\beta(1 - u(t)(V(x,t)-eV(x,0))I(x,t) \\
		f_3(x,u) &= \beta(1 - u(t)) S(x,t) I(x,t)+\beta(1 - u(t)(V(x,t)-eV(x,0))I(x,t) - \gamma I(x,t), \\
		f_4(x,u) &= \gamma I(x,t),
	\end{aligned}
	\]
	\par
	
	\subsection{Development of the Optimal Control}
	
	\noindent $x=[S,V,I,R]^T$	represents the state variables of the system, where \(S\) is the number of susceptible individuals, \(I\) is the number of infected individuals, and \(R\) is the number of recovered individuals. The control function \(u=u(t)\), influences the system by modifying the transmission dynamics, for instance through interventions like vaccination or social distancing.\\
	The set of admissible controls, u is defined as:\\
	\[
	\{\,u \in L^1[0,T]\mid 0 \le u(t) < 1 \bigr\}.
	\]\\
	This means that the control function \(u(t)\) must be a Lebesgue–integrable function over the time interval \([0,T]\), and its value must lie between 0 and 1 for all \(t\in [0,T]\). The lower bound of 0 corresponds to no control being applied, while the upper bound of less than 1 reflects the fact that control measures cannot completely stop transmission but can significantly reduce it \cite{Smirnova2024,Blasioli2023}.\\
	
	\noindent The challenge in this framework is to determine the optimal control \(u(t)\) that minimizes the spread of infection while staying within these constraints, balancing both the effectiveness and feasibility of the interventions.\\
	with an assumed set of initial conditions for the distribution of population. The following constraint is essential.\\
	
	\begin{math}S(x,t)+V(x,t)+ I (x,t) + R(x,t)=1\end{math}\\

	\noindent But as we are considering that the entire population is not totally susceptible, it is likely to use a reproduction number $R_c$ instead of the basic reproduction number  $R_0$. In this case, $R_c$ is estimated as \\
	
	\[
	R_e \;=\; \frac{\beta\bigl(1 - u(t)\bigr)}{\gamma}\,\bigl[S(x,0) + V(x,0)\bigr]
	\;=\; R_0\bigl(u(t)\bigr)\,\bigl[S(x,0) + V(x,0)\bigr].	\]\\
	
	\noindent More over considering the basic reproduction number $R_0 = \beta / \gamma$ , is large, this type of control strategy may not be sustainable for deterministic control functions. A high reproduction number indicates rapid transmission, making it challenging to restrain the spread using early-stage interventions alone. While timely actions such as physical distancing, hygiene measures, and mask-wearing can save lives and protect public health, they also impose significant social, psychological, and economic costs. This creates a difficult balance for policymakers, who must weigh the public health benefits against the adverse side effects of these preventive measures. Mathematically, this leads to solving an optimal control problem where the primary objective is to reduce the daily number of new infections, represented by $\beta (1-u(t))S(t)I(t)$  , while simultaneously minimizing the cost of preventive measures, represented by $\lambda c(u(t))$ . This leads to solving the following objective functional \\
	
	\[
		\min_{u} \, J(\mathbf{x}, u) = \int_{0}^{T} \left[ \beta (1 - u(t)) S(t) I(t) + \lambda \ast c(x(t)) \right] dt
	\]\\
	\noindent From a mathematical standpoint, the solution to this optimal control issue provides the most effective strategy for preventing transmission of disease while minimizing associated costs, especially before vaccination measures can be carried out. The goal is to reduce the entire costs of the epidemic, which include both the public health impact (i.e., fewer infections) and the economic or social costs of the preventive measures implemented.\\
	We aim to demonstrate that the control trajectory u=u(t)  is an optimal control for the objective functional. The functional is constrained by the system $\frac{dx}{dt} = f(x,u)$ with the additional condition that $u(t)\geq 0$ for all $t \in [0,T]$. The cost function   is twice continuously differentiable in its domain containing [0,1), with certain properties ensuring smoothness, positivity, and convexity.Let $\frac{d^2 c}{du^2} >0$  which ensures the convexity of the problem. With the initial cost being a null cost $c(0)=0$. $\frac{dc}{du} >0$ for $u>0$   for  i.e. the cost increases if we increase the value of the control. $\lim_{u \to 1^{-}} c(u) = \infty$ , the cost reaches an infinite value if the maximum value of control is obtained \cite{Smirnova2024,Blasioli2023,Hethcote2000}.

  \subsection{L-BFGS method for adaptive optimal control }
  \label{Effectiveness model} 
  
  \noindent The L-BFGS (Limited memory Broyden-Fletcher-Goldfarb-Shanno) method is a limited-memory quasi-Newton optimization algorithm, particularly suited for large-scale problems. It approximates the Hessian matrix using information from previous iterations, allowing for faster convergence while using significantly less memory than full Hessian methods. This makes L-BFGS ideal for solving the optimal control problem in an epidemic model where the number of variables is large \cite{Carlon2022,Deutz2023}.

 \subsubsection{Discretization of the Objective Functional}
 
 To apply the L-BFGS method, the objective functional J(u)  is discretized over the time interval [0,T] dividing it into N time steps. The discretized form of the functional becomes\\
 \[
 J(u) \approx \sum_{n=0}^{N-1} \left[ \beta (1 - u_n) S_n I_n + \lambda \cdot c(u_n) \right] \Delta t
 \]
 
 \noindent Where $u_n=u(t_n), S_n=S(t_n)$and $I_n=I(t_n)$  are the values in the n-th time step, and $\delta t$ is the time increment.
 
 \subsubsection{Gradient Computation}
 
 \noindent The L-BFGS algorithm requires the computation of the gradient of J(u)  with respect to the control u(t)  The gradient at each time step n is given by:
 \[
 \frac{\partial J}{\partial u_n} = -\beta S_n I_n + \lambda \frac{d}{du_n}c(u_n)
 \]
 
  \subsubsection{Hessian Approximation}
 \noindent The L-BFGS method does not explicitly compute the Hessian matrix. Instead, it approximates the inverse Hessian $H_k^{-1}$  using gradient information from previous iterations. This approximation is used to update the control function $u(t)$  at each iteration. The control update is performed as follows:
 \[
 u_{k+1} = u_k - \alpha_k H_k^{-1} \nabla J(u_k)
 \]
 Where $\alpha_k$  is a step size determined through a line search procedure, and is the gradient at the current iteration \cite {Byrd1995,McAsey2012}.\\
 \noindent The Hessian matrix  $H$  is constructed from the second derivatives of the objective functional. The elements of the Hessian matrix are defined as follows:
 \[
 H_{ij} = \frac{\partial^2 J}{\partial u_i \partial u_j}
 \]
 In our context, let’s assume we have discrete control values at time steps n where i and j represent indices corresponding to different time steps. Thus, the Hessian matrix can be expressed in discrete form:
 \[
 H = \begin{bmatrix}
 	\frac{\partial^2 J}{\partial u_0^2} & \frac{\partial^2 J}{\partial u_0 u_1} & \cdots & \frac{\partial^2 J}{\partial u_0 u_{N-1}} \\
 	\frac{\partial^2 J}{\partial u_1 u_0} & \frac{\partial^2 J}{\partial u_1^2} & \cdots & \frac{\partial^2 J}{\partial u_1 u_{N-1}} \\
 	\vdots & \vdots & \ddots & \vdots \\
 	\frac{\partial^2 J}{\partial u_{N-1} u_0} & \frac{\partial^2 J}{\partial u_{N-1} u_1} & \cdots & \frac{\partial^2 J}{\partial u_{N-1}^2}
 \end{bmatrix}
 \]
 \subsubsection{Computing the Hessian Elements}
 \noindent To compute each element of the Hessian matrix, we need to evaluate the second derivatives of the objective functional J(u)  with respect to $u_n$.\\
 \noindent Using the chain rule and product rule, the second derivative can be expressed as:
 \[
 \frac{\partial^2 J}{\partial u_n^2} = \lambda . \frac{d^2 c(u_n)}{du_n^2}, \text{ for } i = j
 \]
 For the off-diagonal elements 
 \[
 \frac{\partial^2 J}{\partial u_i \partial u_j} = -\beta S_i I_j, \text{ for } i \neq j
 \]
 
  \subsubsection{Final Form of the Hessian Matrix}
  Thus, the Hessian matrix H takes the form:
  \[
  H = \begin{bmatrix}
  	\lambda \cdot \frac{\partial^2 c(u_0)}{\partial u_0^2} & -\beta S_0 I_1 & \cdots & -\beta S_0 I_{N-1} \\
  	-\beta S_1 I_0 & \lambda \cdot \frac{\partial^2 c(u_1)}{\partial u_1^2} & \cdots & -\beta S_1 I_{N-1} \\
  	\vdots & \vdots & \ddots & \vdots \\
  	-\beta S_{N-1} I_0 & -\beta S_{N-1} I_1 & \cdots & \lambda \cdot \frac{\partial^2 c(u_{N-1})}{\partial u_{N-1}^2}
  \end{bmatrix}
  \]
 The Hessian matrix is instrumental in optimizing the control problem. It provides information about the convexity of the objective functional: 
 \begin{itemize}
 	\item If H is positive definite, the objective functional J(u) is locally convex, indicating a local minimum.
 	\item If H is indefinite or negative definite, it suggests that the control strategy might lead to suboptimal results.
 \end{itemize}
 For the Hessian matrix to be positive definite, the diagonal elements $H_{nn}$  must be positive. This requires the second derivative of the cost function $c(u)$ to be positive for all $u \in [0,1)$ \\
 Thus, for each $u_n$ :
\[
\lambda \cdot \frac{\partial^2 c(u_n)}{\partial u_n^2} > 0
\]
Given that $\lambda>0$ and the cost function c(u) is twice continuously differentiable and convex in the interval [0,1), it follows that $\frac{\partial^2 c(u_n)}{\partial u_n^2} > 0$ by assumption. Therefore, the diagonal elements of $H$ are strictly positive. But the off-diagonal elements $H_{ij}=-\beta S_i I_j$ for $i \neq j$ are negative since $S_i, I_j,\beta > 0$. However, negative off-diagonal elements alone do not disqualify the matrix from being positive definite. The overall positive definiteness depends on the structure of the matrix and the relative sizes of the diagonal elements.\\
\noindent To check whether the Hessian matrix H is positive definite, we examine the eigenvalues of H. A matrix is positive definite if and only if all its eigenvalues are positive. Since the diagonal elements are strictly positive, and the off-diagonal elements are bounded by the product $-\beta S_i I_j$.we can use Gershgorin’s Circle Theorem to evaluate the eigenvalues. For each diagonal element we define a Gershgorin disk centered at $H_{nn}$ with radius\\
\[
R_n = \sum_{j \neq n} \left| H_{nj} \right|
\]
The eigenvalues of the matrix H are guaranteed to lie within these Gershgorin disks. Thus, for positive definiteness, the following condition must hold:
\[
\lambda \cdot \frac{\partial^2 c(u_n)}{\partial u_n^2} > \sum_{j=n} |\beta S_n I_j|
\]
This means that the diagonal elements must dominate the sum of the magnitudes of the off-diagonal elements. Given the structure of the model, with a properly chosen $\lambda$  and cost function c(u) , this condition can be satisfied, ensuring that all eigenvalues are positive. Hence this guarantees that the optimal control found by the L-BFGS method will indeed minimize the functional effectively \cite {Carlon2022}.

  \subsubsection{Forward-Backward Sweep Method}
  To solve the optimal control problem, the forward-backward sweep method is employed:
   \begin{itemize}
  	\item Forward Sweep: The state equations (SIR-V model) are solved forward in time using the current control $u_k(t)$
  	\item Backward Sweep: The adjoint equations are solved backward in time to compute the gradient $\nabla J(u_k)$  , which is used to update the control \cite {McAsey2012,Sharp2021}.
  \end{itemize}

\subsection{Payoff Structure}

\label{Payoff Structure}

An epidemic season continues until all infected individuals have recovered eventually. During this time, unvaccinated individuals who become infected bear a cost of infection, denoted as $C_i$. In contrast, unvaccinated individuals who manage to avoid infection face no cost.\\

\noindent For those who take preventive measures—whether through vaccination or other defenses—but still become infected, they incur a combined cost. This combined cost includes both the cost of vaccination $C_v$  and the cost of infection $C_i$ , leading to a total cost of $C_v+C_i$\\

\noindent To simplify the analysis, a relative cost of vaccination is introduced, represented as $C_r$, where $C_r$ is ratio of the vaccination cost to the infection cost $C_r=C_v/C_i$, with $C_i$ normalized to 1 and $C_r$ ranging between 0 and 1. This normalization helps in comparing different outcomes without losing generality. Consequently, the payoff each individual receives by the end of the epidemic season depends on their final state whether they remain healthy or get infected and whether they chose a preventive measure, such as vaccination. To answer the question about the implementation of this strategy ensures that

\begin{table}[h!]
	\centering
	
	\caption{Payoff structure projections at the end of each epidemic season.}
	
	\begin{tabular}{|c|c|c|}
		\hline
		\textbf{State} & \textbf{Actione} & \textbf{Payoff} \\ \hline
	
	\textbf{Healthy }     & No vaccination      & $0$  \\ \hline

	\textbf{Healthy }     & Vaccination   & $C_{v}$ \\ \hline
	
	\textbf{Infected }   & No vaccination    & $C_i$ \\ \hline
	
	\textbf{Infected }   & Vaccination    & $C_v+C_i$  \\ \hline
	
\end{tabular}

\label{table:2}
\end{table}

\noindent We may now analyze the predicted payoffs in terms of three categories: the average societal payoff $\langle \pi \rangle$, the average payoff for cooperators (vaccinated people) $\langle \pi_C \rangle$, and the average payoff for defectors (unvaccinated individuals) $\langle \pi_D \rangle$. These categories correspond to the many tactics that people adopt, such as incomplete vaccination or other contagion resistance strategies.\\
(For an effectiveness model)\\
\begin{align}
	\langle \pi \rangle &= -C_r x (e+(1 - e) \exp[-R_0 R(x, \infty)]) - (C_r + 1) x(1 - e)(1 - \exp[-R_0 R(x, \infty)]) \nonumber \\
	&\quad - (1 - x) [1 - \exp[-R_0 R(x, \infty))] \\
	\langle \pi_C \rangle &= -C_r (e + (1 - e) \exp[-R_0 R(x, \infty)]) - (C_r + 1) (1 - e)(1 - \exp[-R_0 R(x, \infty)]) \\
	\langle \pi_D \rangle &= -(1 - \exp[-R_0 R(x, \infty)])
\end{align}

\subsection{Strategy adaptation}
\noindent In the context of the vaccination game, individuals are given the opportunity to revise their strategies at the conclusion of an epidemic season, deciding whether to use a provision based on the outcomes of the previous season. This study examines three types of strategy updates introduced in earlier research. These previous studies employed a multi-agent simulation (MAS) approach, incorporating a spatial structure among individuals through an underlying network that connected agents. In contrast, the current research does not account for spatial structure, instead utilizing the mean field approximation to assess a neighbor's payoff.

\subsubsection{Individual-based risk assessment (IB-RA)}
\label{Individual-based risk assessment (IB-RA)}
Individual-based risk assessment (IB-RA) is a framework where individuals adjust their strategies by observing and assessing the behavior of those around them. This approach mirrors the well-established pairwise Fermi updating rule, which is commonly used in stochastic games involving two players and two possible strategies (2 × 2 games). Under this rule, an individual randomly selects a neighboring player and evaluates their strategy based on the perceived success or payoff of that neighbor. The decision to adopt the neighbor's strategy is not automatic but probabilistic, meaning individual i will adopt the strategy of the selected neighbor j with a certain probability, influenced by how advantageous or successful the neighbor’s strategy appears. This allows individuals to continuously refine their strategies over time based on local interactions and information, promoting adaptive decision-making in dynamic environments.
\[
P(s_i \leftarrow s_j) = \frac{1}{1 + \exp[-(\pi_j - \pi_i) / K]}
\]

\noindent In this context, $S_i$  represents the strategy employed by individual i, while $\pi_i$ is the payoff received by i during the previous season. The parameter $\kappa $(greater than 0) captures the intensity of selection, indicating how responsive individuals are to differences in payoffs. A smaller value of $\kappa $ implies greater sensitivity to variations in payoffs, meaning individuals are more likely to adjust their strategy based on even slight differences. In this study, we set $\kappa $ to 0.1. This strategy-updating method is referred to as individual-based risk assessment (IB-RA).\\

\noindent Within the current framework, there are four distinct categories of individuals based on cost burden: (i) a successful free-rider (SFR) who pays nothing, (ii) a failed free-rider (FFR) who incurs a cost of -1, (iii) an infected vaccinator (IV) who incurs a cost of $C_v-1$ , and (iv) a healthy vaccinator (HV) who bears the cost of $-C_v$ . Individuals can adopt one of two strategies: using vaccination (referred to as V) or not using vaccination (referred to as NV). Consequently, the probability governing the transition time for x, which must be factored into the IB-RA rule, falls into one of the following eight possible cases:

\begin{equation}
	P(HV \leftarrow SFR) = \frac{1}{1 + \exp\left[-\frac{0 - (-C_r)}{\kappa}\right]}
	\label{eq:prob1}
\end{equation}

\begin{equation}
	P(HV \leftarrow FFR) = \frac{1}{1 + \exp\left[-\frac{-1 - (-C_r)}{\kappa}\right]}
	\label{eq:prob2}
\end{equation}

\begin{equation}
	P(IV \leftarrow SFR) = \frac{1}{1 + \exp\left[-\frac{0 - (-C_r-1)}{\kappa}\right]}
	\label{eq:prob3}
\end{equation}

\begin{equation}
	P(IV \leftarrow FFR) = \frac{1}{1 + \exp\left[-\frac{-1 - (-C_r - 1)}{\kappa}\right]}
	\label{eq:prob4}
\end{equation}

\begin{equation}
	P(SFR \leftarrow HV) = \frac{1}{1 + \exp\left[-\frac{-C_r - 0}{\kappa}\right]}
	\label{eq:prob5}
\end{equation}

\begin{equation}
	P(SFR \leftarrow IV) = \frac{1}{1 + \exp\left[-\frac{-C_r - 1 - 0}{\kappa}\right]}
	\label{eq:prob17}
\end{equation}

\begin{equation}
	P(FFR \leftarrow HV) = \frac{1}{1 + \exp\left[-\frac{-C_r + 1 }{\kappa}\right]}
	\label{eq:prob7}
\end{equation}

\begin{equation}
	P(FFR \leftarrow IV) = \frac{1}{1 + \exp\left[-\frac{-C_r - 1 + 1 }{\kappa}\right]}
	\label{eq:prob8}
\end{equation}

\subsubsection{Strategy-based risk assessment (SB-RA)}

\noindent Strategy-based risk assessment (SB-RA) modifies the imitation probability used in individual-based risk assessment (IB-RA) to account for situations where individuals evaluate risks based on a socially averaged payoff. This adjustment reflects the influence of widespread information about epidemics, often disseminated by the media. The revised probability is
\[
P(s_i \leftarrow s_j) = \frac{1}{1 + \exp[-(\langle \pi_j \rangle-\langle \pi_i \rangle / K]}
\]
\noindent In this context, $\langle \pi_j \rangle$ represents the average payoff calculated by averaging the collective payoffs of individuals who adopt the same strategy as a randomly selected neighbor j of individual i. This approach allows individuals to assess their potential outcomes not just based on their own experiences but by considering the performance of others who employ similar strategies. By doing so, individuals can make more informed decisions about whether to imitate their neighbors. This updating mechanism is referred to as strategy-based risk assessment (SB-RA), as it incorporates social dynamics and the influence of collective behavior into the decision-making process(SB-RA).\\

\noindent The transition probability that we need to consider in this framework encompasses one of the following possibilities. Each of these scenarios will influence how individuals shift between strategies based on their assessments of payoffs and social interactions:

\begin{equation}
	P(HV \leftarrow NV) = \frac{1}{1 + \exp\left[-\frac{\langle \pi_D \rangle - (-C_r)}{\kappa}\right]}
\end{equation}

\begin{equation}
	P(IV \leftarrow NV) = \frac{1}{1 + \exp\left[-\frac{\langle \pi_D \rangle - (-C_r-1)}{\kappa}\right]}
\end{equation}

\begin{equation}
	P(SFR \leftarrow V) = \frac{1}{1 + \exp\left[-\frac{\langle \pi_c \rangle - 0}{\kappa}\right]}
\end{equation}

\begin{equation}
	P(FFR\leftarrow V) = \frac{1}{1 + \exp\left[-\frac{\langle \pi_C\rangle - (-1)}{\kappa}\right]}
\end{equation}

\subsubsection{Direct commitment (DC)}

\noindent Direct commitment (DC) introduces a new updating rule that defines an individual’s probability of using vaccination, denoted as $P_{S=V}$. This probability is influenced by the individual's awareness of the risks associated with trying to be a free-rider. Specifically, it is assessed by comparing the average payoffs of individuals who use vaccination versus those who do not, as observed within the individual’s neighborhood. This approach allows individuals to make informed decisions about vaccination based on the collective experiences of their peers, taking into account the potential dangers of free-riding behavior.
\[
P_{s=V} = \frac{1}{1 + \exp\left( -\frac{\left( \langle \pi_{s=V} \rangle - \langle \pi_{s,NV} \rangle \right)}{\kappa} \right)}, \quad P_{s=NV} = 1 - P_{s=V}
\]

\noindent $\pi_{s=V}$ and $\pi_{s,NV}$  represent the average payoffs for individuals who use vaccination and those who do not, respectively, as observed within the agent's neighborhood. When considered independently, $P_{s=NV}$ indicates the probability of an individual choosing not to use vaccination. In the present model, the mean field approximation allows for these two probabilities to be represented in terms of $\langle \pi_{C} \rangle$ and $\langle \pi_{D} \rangle$, which reflect the respective payoffs associated with each choice. This updating mechanism is referred to as direct commitment (DC), as it emphasizes the decision-making process based on the perceived risks and benefits of vaccination in relation to the behaviors of others in the community.

\begin{equation}
	P(V \leftarrow NV) = \frac{1}{1 + \exp\left[-\frac{\langle \pi_D \rangle - \langle \pi_C \rangle}{\kappa}\right]}
\end{equation}

\begin{equation}
	P(V \leftarrow NV) = \frac{1}{1 + \exp\left[-\frac{\langle \pi_C \rangle - \langle \pi_D \rangle}{\kappa}\right]}
\end{equation}

\noindent After each epidemic season, individuals update their strategies based on the outcomes, which leads to changes in the proportion of individuals choosing a particular strategy, represented by x.There are two distinct epidemic models: the effectiveness model and the efficiency model. Additionally, there are three different strategy updating rules: IB-RA (Imitation-Based with Random Aspirations), BB-RA (Best-Response with Random Aspirations), and DC (Direct Comparison). This gives rise to six distinct types of dynamics, each resulting from the combination of one epidemic model with one updating rule. These dynamics reflect how individuals adjust their behavior in response to the results of the epidemic season, either by adopting more effective strategies or by modifying their approach to maximize personal payoff.

\subsubsection{Effectiveness model IB-RA}

\begin{equation*}
	\begin{aligned}
		\frac{dx}{dt} &= x (1 - x) \left( e + (1 - e) \exp[-R_0 R(x, \infty)] \right) \exp[-R_0 R(x, \infty)] \left( P(SFR \leftarrow HV) - P(HV \leftarrow SFR) \right) \\
		&+ x (1 - x) \left( e + (1 - e) \exp[-R_0 R(x, \infty)] \right) \left( 1 - \exp[-R_0 R(x, \infty)] \right) \left( P(FFR \leftarrow HV) - P(HV \leftarrow FFR) \right) \\
		&+ x (1 - x) (1 - e) \left( 1 - \exp[-R_0 R(x, \infty)] \right) \exp[-R_0 R(x, \infty)] \left( P(SFR \leftarrow IV) - P(IV \leftarrow SFR) \right) \\
		&+ x (1 - x) (1 - e) \left( 1 - \exp[-R_0 R(x, \infty)] \right) \left( P(FFR \leftarrow IV) - P(IV \leftarrow FFR) \right).
	\end{aligned}
\end{equation*}

\subsubsection{Effectiveness model SB-RA}

\begin{equation*}
  \begin{aligned}
	\frac{dx}{dt} &= -x(1 - x) \left( e + (1 - e) \exp[-R_0 R(x, \infty)] \right) P(HV \leftarrow NV) \\
	&- x(1 - x) (1 - e) \left( 1 - \exp[-R_0 R(x, \infty)] \right) P(IV \leftarrow NV) \\
	&+ x(1 - x) \exp[-R_0 R(x, \infty)] P(SFR \leftarrow V) \\
	&+ x(1 - x) \left( 1 - \exp[-R_0 R(x, \infty)] \right) P(FFR \leftarrow V)
  \end{aligned}
\end{equation*}

\subsubsection{Effectiveness model DC}

\begin{equation*}
  \begin{aligned}
		\frac{dx}{dt} &= -xP(V \leftarrow NV)+(1-x)P(NV \leftarrow V)
  \end{aligned}
\end{equation*}
		
\noindent It is worth noting that Effectiveness model DC equation aligns qualitatively with replicator dynamics, a fundamental concept in evolutionary game theory used to describe the dynamics of systems over time. This connection is important because replicator dynamics is one of the most widely-used models for understanding how strategies evolve within populations.\\
\noindent All the previously discussed dynamical equations can be solved numerically. To achieve this, an explicit scheme is introduced to handle the time-varying components, allowing us to find a numerical solution. This approach enables the calculation of vaccination coverage at equilibrium, showing how the population's behavior stabilizes over time based on the dynamics of vaccination and infection.\\

\noindent The FES and other fractions related to it are given by
\begin{equation*}
\begin{aligned}
	S(x, \infty) &= (1 - x) \exp[-R_0 R(x, \infty)] \\
	V(x, \infty) &= x \exp[-(1 - \eta) R_0 R(x, \infty)] \\
	R(x, \infty) &= (1 - ex) (1 - \exp[-R_0 R(x, \infty)])
\end{aligned}
\end{equation*}

\section {Result and discussion}
\label{Result and discussion}

\begin{figure}[H]
	\centering
	\includegraphics[width=\textwidth, scale=0.8]{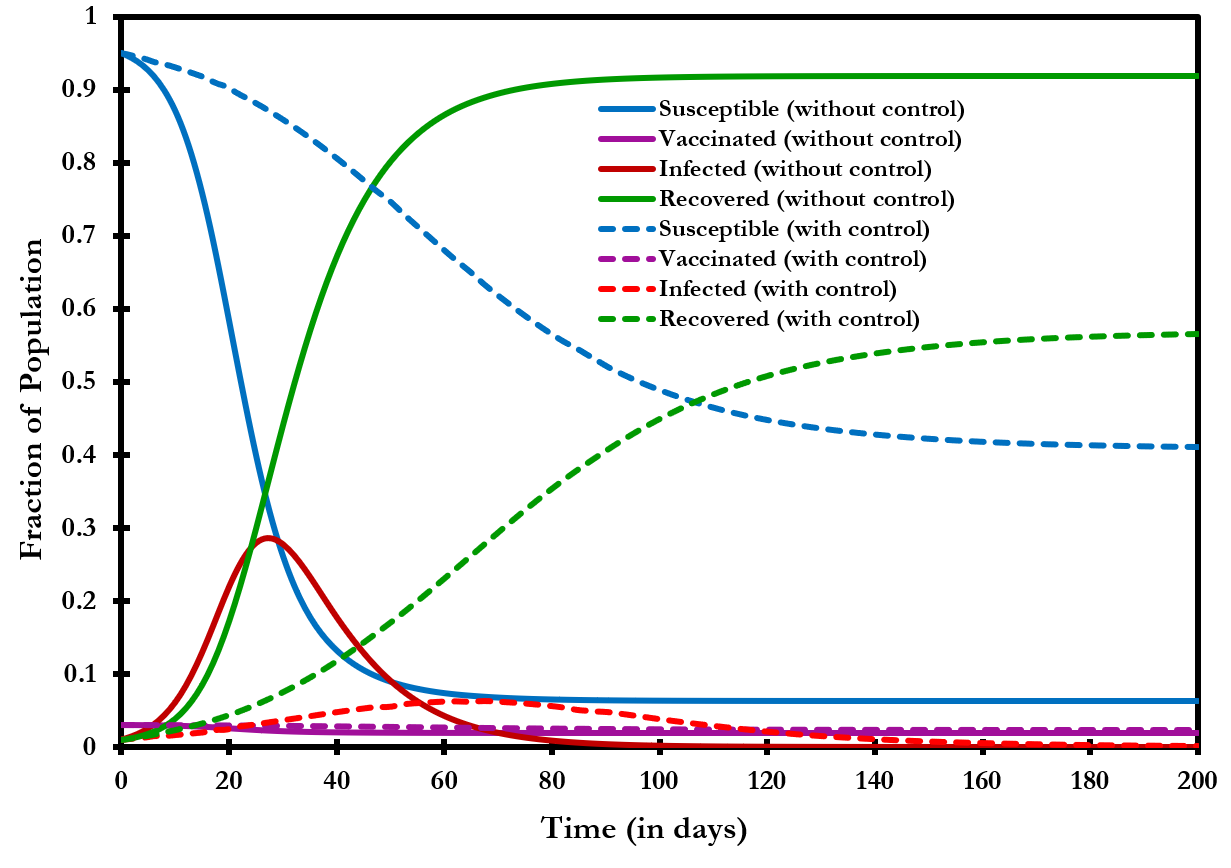}
	\caption{Schematic diagram of SIR-V model with and without control}
	\label{fig:1}
\end{figure}

\noindent In the absence of control, the susceptible population rapidly declines as infections surge, resulting in a sharp infection peak around earlier days \cite {Kirschner1997}\cite{Bertozzi2020}. Conversely, with control, the decline in susceptibility is slower, as vaccination efforts reduce the spread, leading to a flattened and delayed infection curve as depicted by the schematic of Fig. \ref{fig:1}. The vaccinated population grows significantly under control measures, illustrating the effectiveness of practical interventions in mitigating the epidemic. The recovered population also shows slower growth under control, as fewer individuals require natural immunity due to reduced infections. This highlights the shift from reliance on recovery to achieving immunity via vaccination as shown in Fig. \ref{fig:1}. Control strategies demonstrate a substantial reduction in the infection peak, easing the burden on healthcare systems and providing critical time for preparedness. Additionally, the prolonged decline in susceptibility reflects sustained efforts to manage the disease over time \cite {Smirnova2024}. The results emphasize the importance of implementing adaptive, optimal control measures, particularly vaccination, to prevent widespread infections and protect susceptible populations.

\begin{figure}[H]
	\centering
	\includegraphics[width=\textwidth]{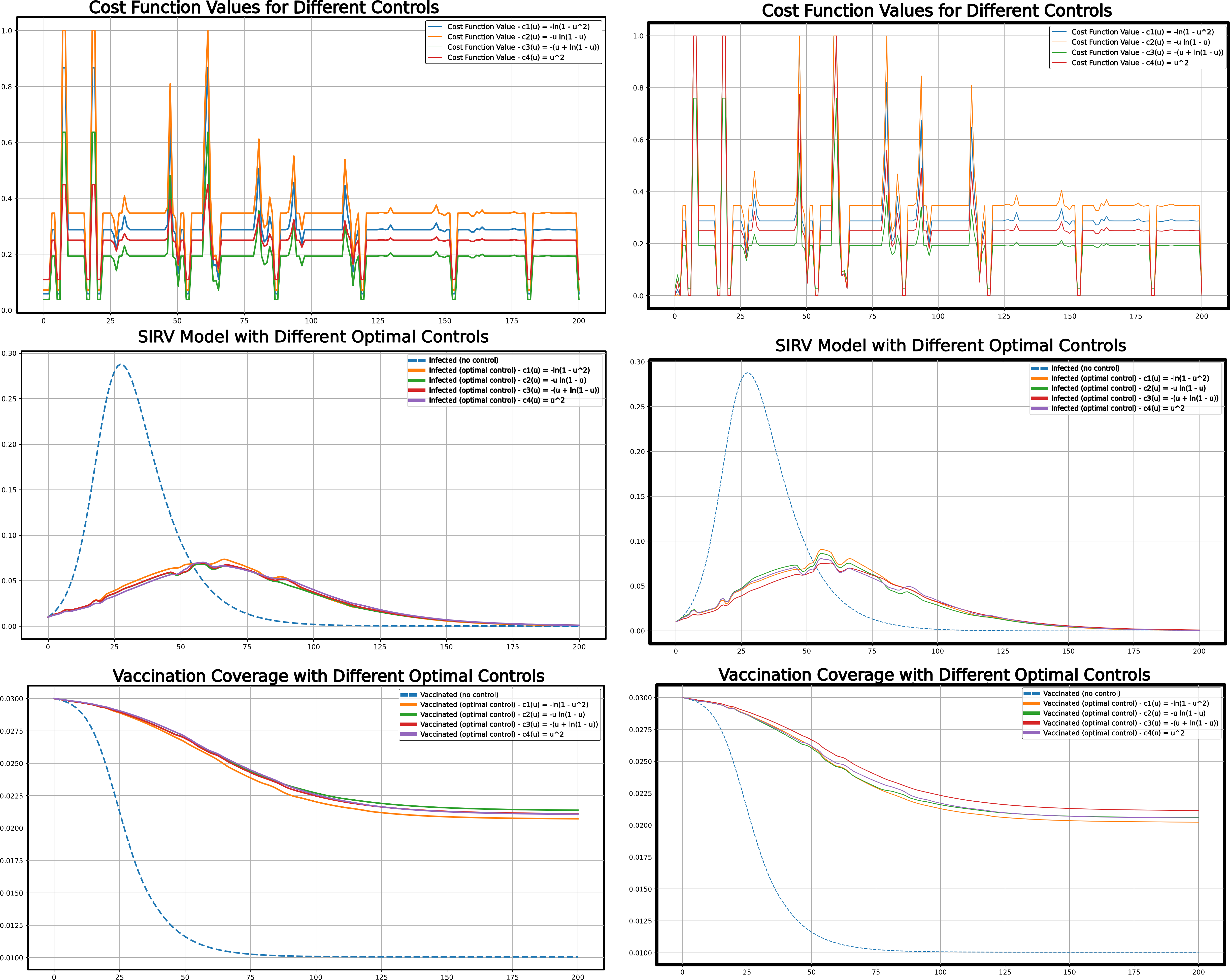}
	\caption{(a) Adaptive control cost function (b)Infected Population with and without control (c) Vaccinated Population with and without control}
	\label{fig:2}
\end{figure}

\noindent The Fig. \ref{fig:2} illustrates the impact of adaptive cost functions on optimal control strategies for managing an epidemic in the SIR-V model under two scenarios: $\lambda= 1 $ and $\lambda= 0.1 $. The cost function plots show that for $\lambda= 1 $, the control measures are more dynamic, reflecting a higher emphasis on minimizing infections despite increased vaccination costs. In contrast, for $\lambda= 0.1 $, the cost functions are smoother, indicating steadier and less aggressive interventions, as lower $\lambda$ prioritizes cost efficiency over rapid infection reduction. The infected population plots highlight that optimal control significantly reduces and flattens the infection peak compared to the no-control scenario. For $\lambda= 1 $, the infection reduction is more noticeable, demonstrating the effectiveness of adaptive control strategies in mitigating disease spread. Meanwhile, for $\lambda= 0.1 $, the infection reduction is slightly less intense but still achieves a considerable decline. The vaccinated population increases under all control strategies, with faster vaccination for $\lambda= 1 $ due to more aggressive interventions. For $\lambda= 0.1 $, vaccination occurs more gradually but consistently. These findings underscore the importance of tuning cost functions and $\lambda$ to balance infection mitigation, vaccination effort, and resource efficiency for effective epidemic control \cite{Smirnova2024,Blasioli2023}. The cost function for the preventive measure, c(u(t)), exhibits oscillatory behavior initially requiring intensive interventions during the early stage of the epidemic, followed by a reduction in effort as the situation stabilizes. This strategy helps lower the disease transmission rate while also reducing costs, since interventions are applied primarily when the transmission rate begins to rise.

\begin{figure}[H]
	\centering
	\includegraphics[width=\textwidth]{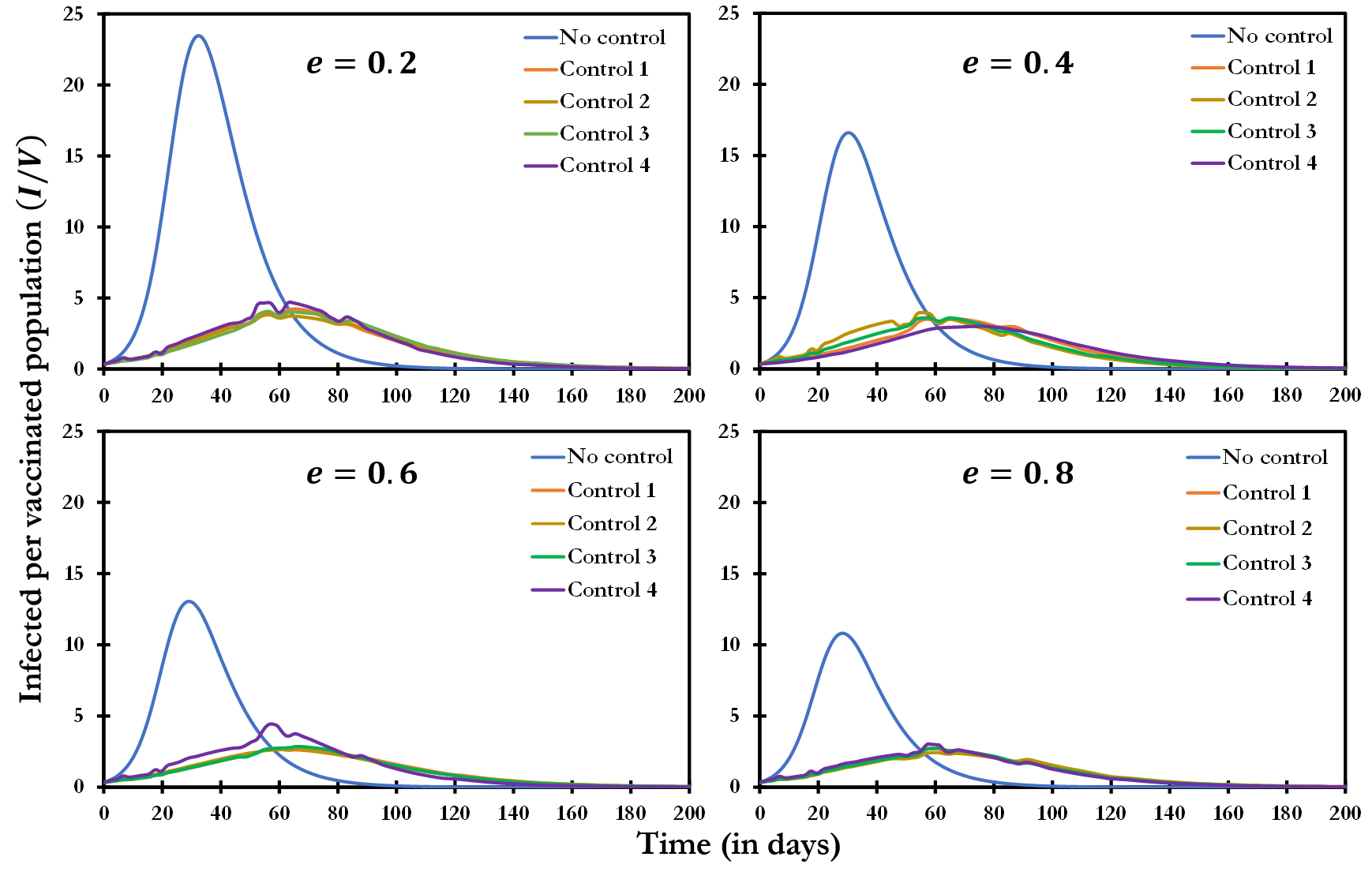}
	\caption{Infected population per vaccinated population ($I/V$) with variations of effectiveness with/ without control}
	\label{fig:3}
\end{figure}

\noindent The infected population per vaccinated population (\( I/V \)) for varying vaccine effectiveness values (\( e = 0.2,\, 0.4,\, 0.6,\, 0.8 \)) under different control cost functions and a no-control scenario is shown in \textbf{Fig. \ref{fig:3}}. As vaccine effectiveness improves, the infected population decreases significantly across all scenarios, emphasizing the critical role of high vaccine efficacy in mitigating disease spread. The no-control scenario consistently shows the highest infection peaks and \( I/V \) ratios, underscoring the necessity of implementing control measures, especially for lower vaccine effectiveness values (\( e = 0.2,\, 0.4 \)). Among the cost functions, \( c(u) = -\ln(1 - u^2) \) demonstrates better performance in reducing the \( I/V \) ratio when vaccine effectiveness is low (\( e = 0.2,\, 0.4 \)), as it balances infection reduction with resource efficiency. For higher vaccine effectiveness (\( e = 0.6,\, 0.8 \)), the differences between cost functions diminish, suggesting that all control strategies achieve similar outcomes~\cite{Kirschner1997}. This indicates that the choice of cost function becomes more critical when vaccine effectiveness is low, whereas higher vaccine efficacy allows for greater flexibility in selecting cost functions. Overall, the logarithmic cost function \( c(u) = -\ln(1 - u^2) \) is recommended for scenarios with lower vaccine efficacy due to its ability to efficiently reduce infections while balancing control efforts. For higher vaccine effectiveness, simpler cost functions like \( c(u) = u^2 \) is also effective, achieving similar reductions in infections with less complexity~\cite{Smirnova2024}. These findings highlight the importance of tailoring cost strategies to vaccine efficacy, ensuring optimal resource utilization while minimizing the disease burden~\cite{Hethcote2000}.

\begin{figure}[H]
 	\centering
 	\includegraphics[width=\textwidth]{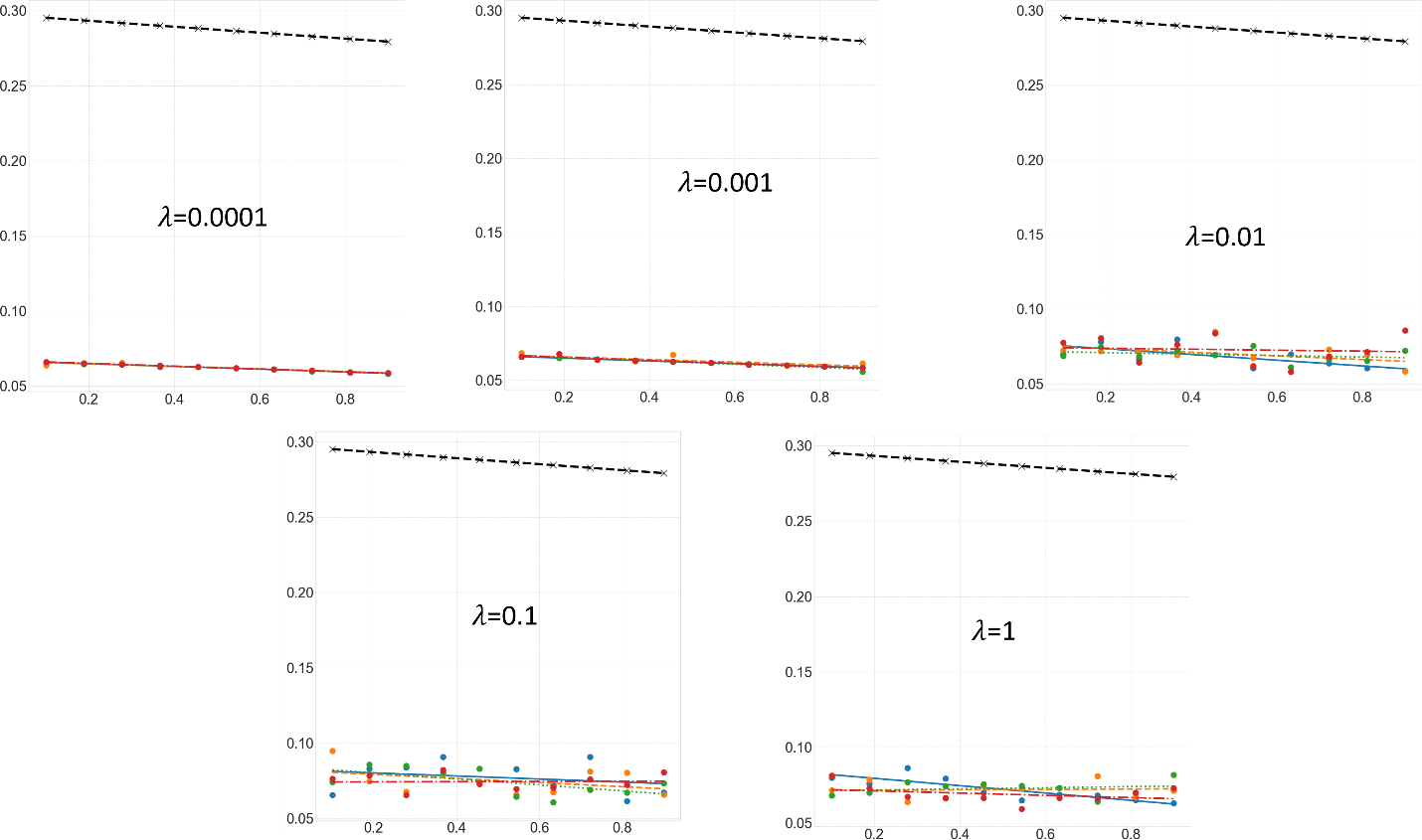}
 	\caption{Maximum infected population with increasing effectiveness considering varying regularization parameter.}
 	\label{fig:4}
 \end{figure}

 \noindent \textbf{Fig. \ref{fig:4}} illustrates the stability of maximum infected individuals (\( I_{\text{max}} \)) versus effectiveness (\( e \)) under varying regularization parameters (\( \lambda \)) for both controlled and uncontrolled cases. Mathematically, regularization introduces a penalty term in the optimization process to prevent overfitting, stabilizing the model's predictions. For small \( \lambda \) values (e.g., \( \lambda = 0.0001 \)), the controlled curves converge further along higher effectiveness of vaccines, demonstrating reduced sensitivity to small perturbations in control parameters~\cite{Smirnova2024}. This aligns with the regularization principle that higher \( \lambda \) values smoothen variations by penalizing complex control strategies (\( \|u\|^2 \)). Across all \( \lambda \) values, \( I_{\text{max}} \) decreases monotonically with increasing \( e \), showcasing the direct proportionality between effectiveness and disease mitigation. The uncontrolled case, which lacks regularization or intervention, consistently has higher \( I_{\text{max}} \) values, independent of \( \lambda \), reaffirming the necessity of control strategies. Stability analysis reveals that moderate regularization (\( 0.001 \leq \lambda \leq 0.01 \)) balances performance and robustness, with controlled cases maintaining consistently lower \( I_{\text{max}} \) and narrower variances. This demonstrates that introducing controls governed by a well-chosen \( \lambda \) can effectively stabilize disease spread while mitigating risks of overly complex or simplistic strategies~\cite{Lenhart2007,Bock1984}.
 
\begin{figure}[H]
	\centering
	\includegraphics[width=\textwidth]{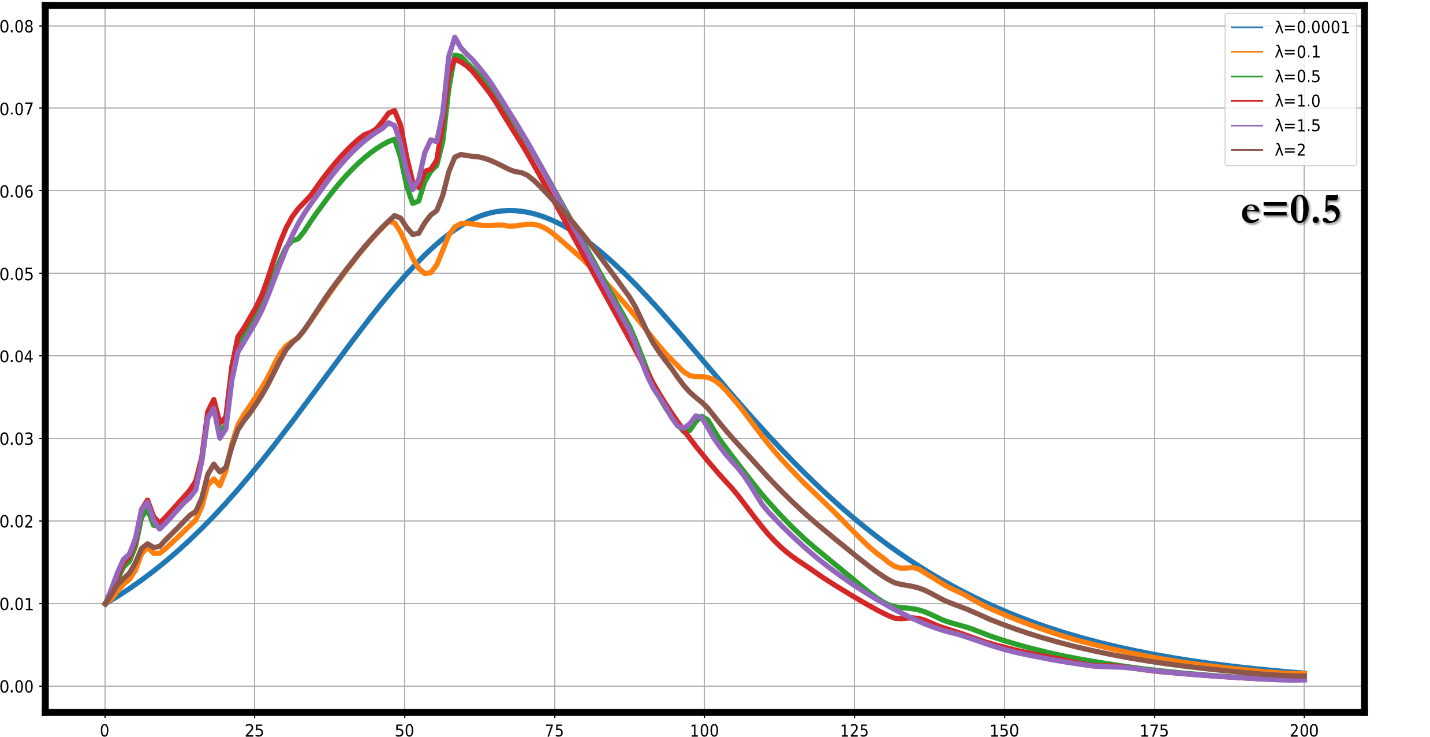}
	\caption{Infected population trend with time considering varying regularization parameter.}
	\label{fig:5}
\end{figure}

 \noindent The trend of infected population varies with the regularization parameter $\lambda$ at a fixed effectiveness \( e = 0.5 \), aligning with earlier analyses. For small lambda values (e.g., \( \lambda = 0.0001 \)), the infection curve peaks higher and later, indicating more aggressive but unstable control strategies. As lambda increases, the peaks lower and stabilize, with smoother dynamics and faster declines in infections, reflecting the trade-off between control complexity and stability. Moderate lambda values (e.g., \( \lambda = 0.5, 1 \)) in \textbf{Fig. \ref{fig:5}} achieve a balance, reducing peak infections while maintaining system stability, reinforcing the importance of fine-tuning $\lambda$ in control optimization~\cite{Lenhart2007}.
 
\begin{figure}[H]
 	\centering
 	\includegraphics[width=\textwidth]{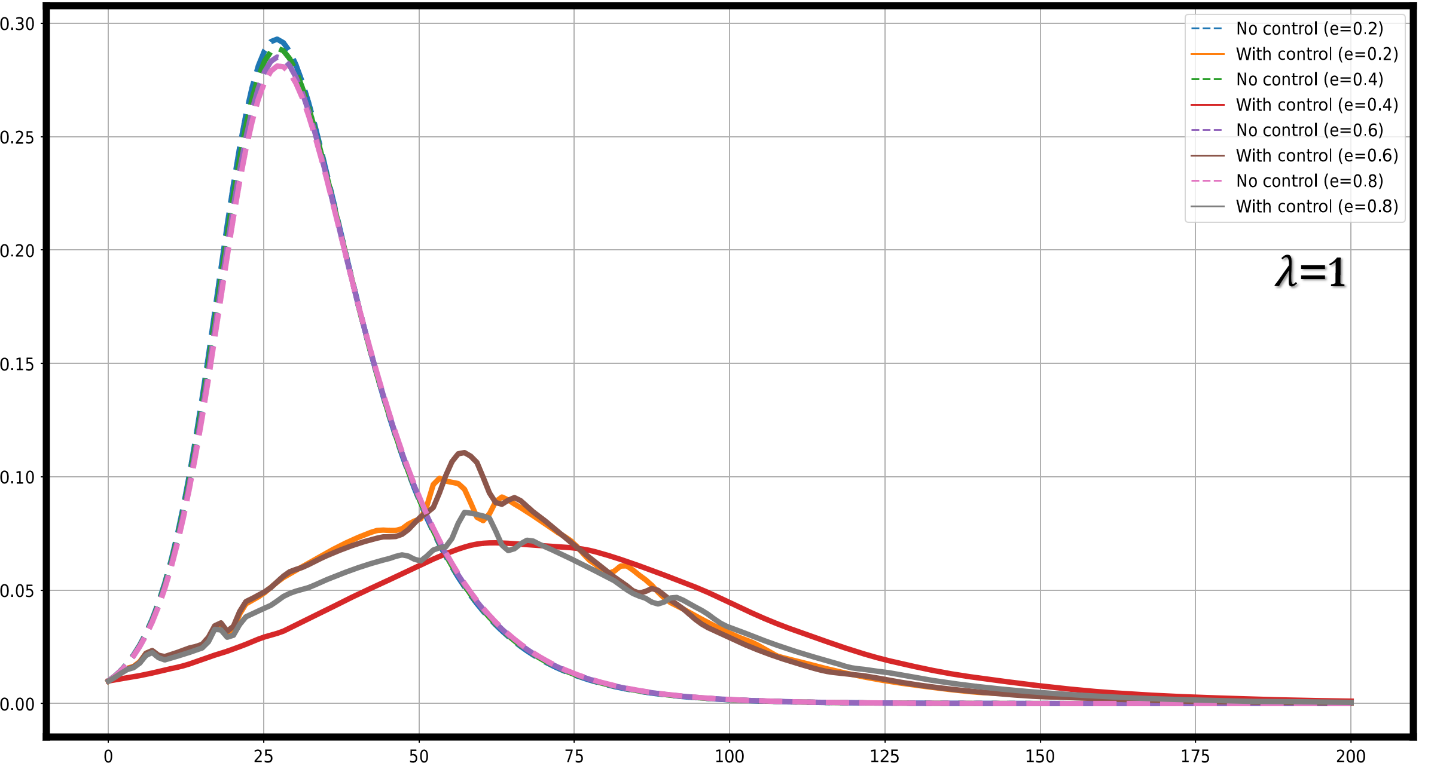}
 	\caption{Infected population trend with time considering varying vaccination effectiveness with/without control.}
 	\label{fig:6}
 \end{figure}

 \noindent In the uncontrolled case, the infection curve follows a predictable pattern: rapid growth, a sharp peak, and subsequent decline, irrespective of effectiveness levels. This consistency arises because no interventions modify the natural progression of the disease, which is primarily driven by the basic reproduction number (\( R_0 \)). For \( R_0 > 1 \), the disease spreads rapidly until natural factors, like immunity or resource depletion, curb its growth in \textbf{Fig. \ref{fig:6}}.\\
 
 \noindent In contrast, the controlled scenario shows significant variation in infection dynamics as effectiveness increases. Higher effectiveness leads to lower infection peaks and faster reductions in infections, as interventions effectively disrupt disease transmission~\cite{Hethcote2000,Lenhart2007,Huang2022}. The impact of effectiveness is amplified in the controlled case because it directly modifies \( R_0 \), lowering it toward or below the critical threshold of 1, which slows or halts the epidemic. When \( R_0 \) is reduced closer to 1, the disease spread stabilizes, leading to smaller outbreaks and quicker resolution. Conversely, for lower effectiveness, the control measures are less effective, causing outcomes that resemble the uncontrolled case~\cite{Bock1984}.
 
\begin{figure}[H]
 	\centering
 	\includegraphics[width=\textwidth]{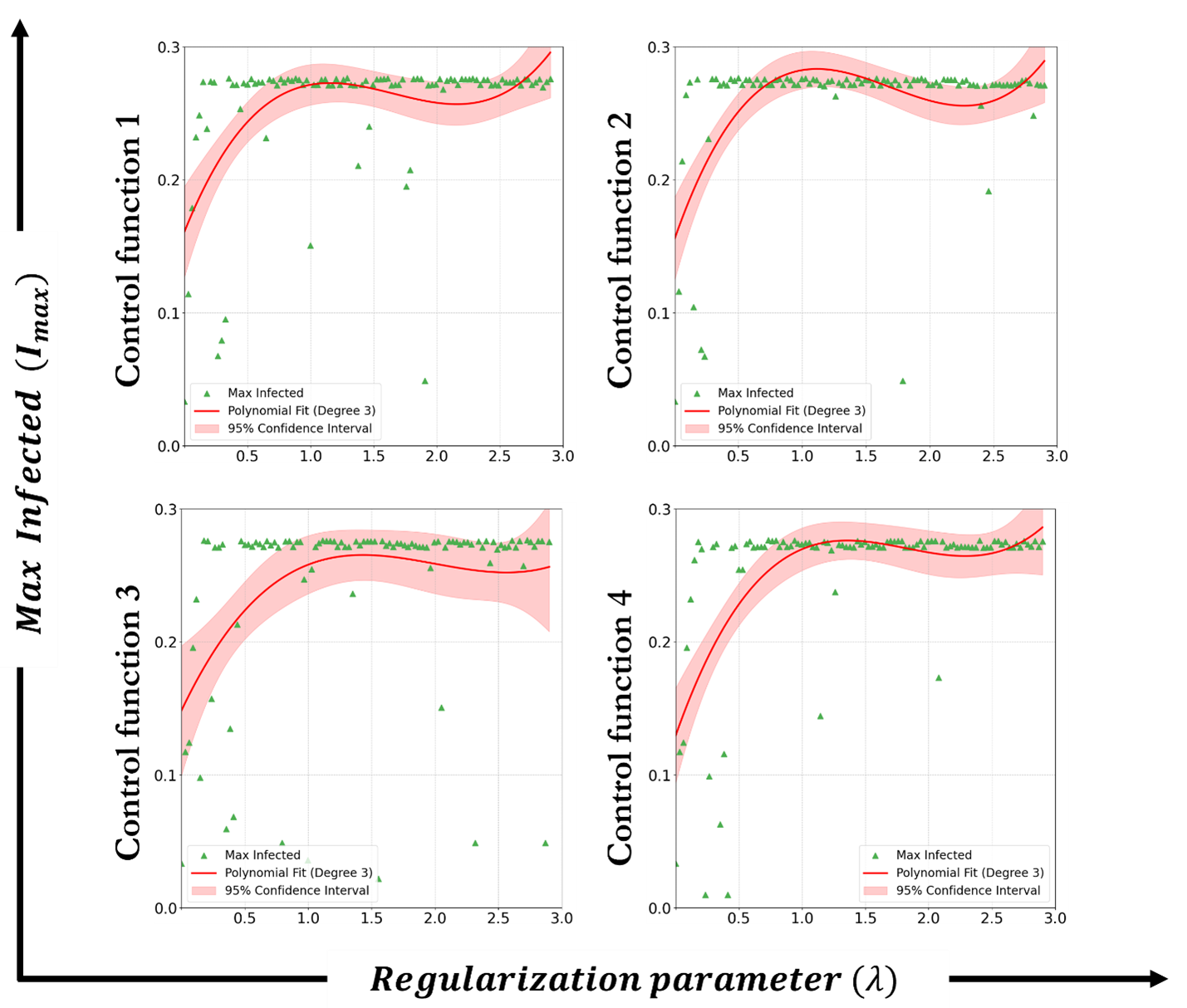}
 	\caption{Maximum Infected population trend with increasing regularization parameter.}
 	\label{fig:7}
 \end{figure}

  \noindent \textbf{Fig. \ref{fig:7}} analysis shows that the peak infected population has variations with changing regularization parameter. For the uncontrolled case, the peak infected population has comparatively stable trends, indicating the absence of control effects. Consequently, the natural dynamics of the epidemic prevail, and regularization has little to no effect.\\
	
  \noindent In the controlled case, nevertheless, the peak infected population is more sensitive to the variation of the regularization parameter. The sensitivity is due to the dependence of control policies on regularization, and the stability of the system is affected directly. With increased regularization, there are constraints imposed on the efficiency of control interventions. When optimized, the peak infection is decreased considerably, a sign of enhanced stability \cite{Diekmann2010,Kirk2004}. Excessive regularization, however, may cause over-constrained dynamics, lowering control efficacy and increasing infection peaks.\\
	
  \noindent The oscillatory behavior in the controlled plots emphasizes the interaction between control strategies and epidemic dynamics. At low values of regularization, interventions are still effective as they restrict infections and provide stability. After a point, the regularization starts to reduce the flexibility of control measures, and higher maximum infections result. This indicates that although regularization is needed to stabilize the control system, it must be fine-tuned to achieve a trade-off between effectiveness and prevention of inefficiencies.\cite{Lenhart2007,Bock1984}.
  
\begin{figure}[H]
	\centering
	\includegraphics[width=\textwidth]{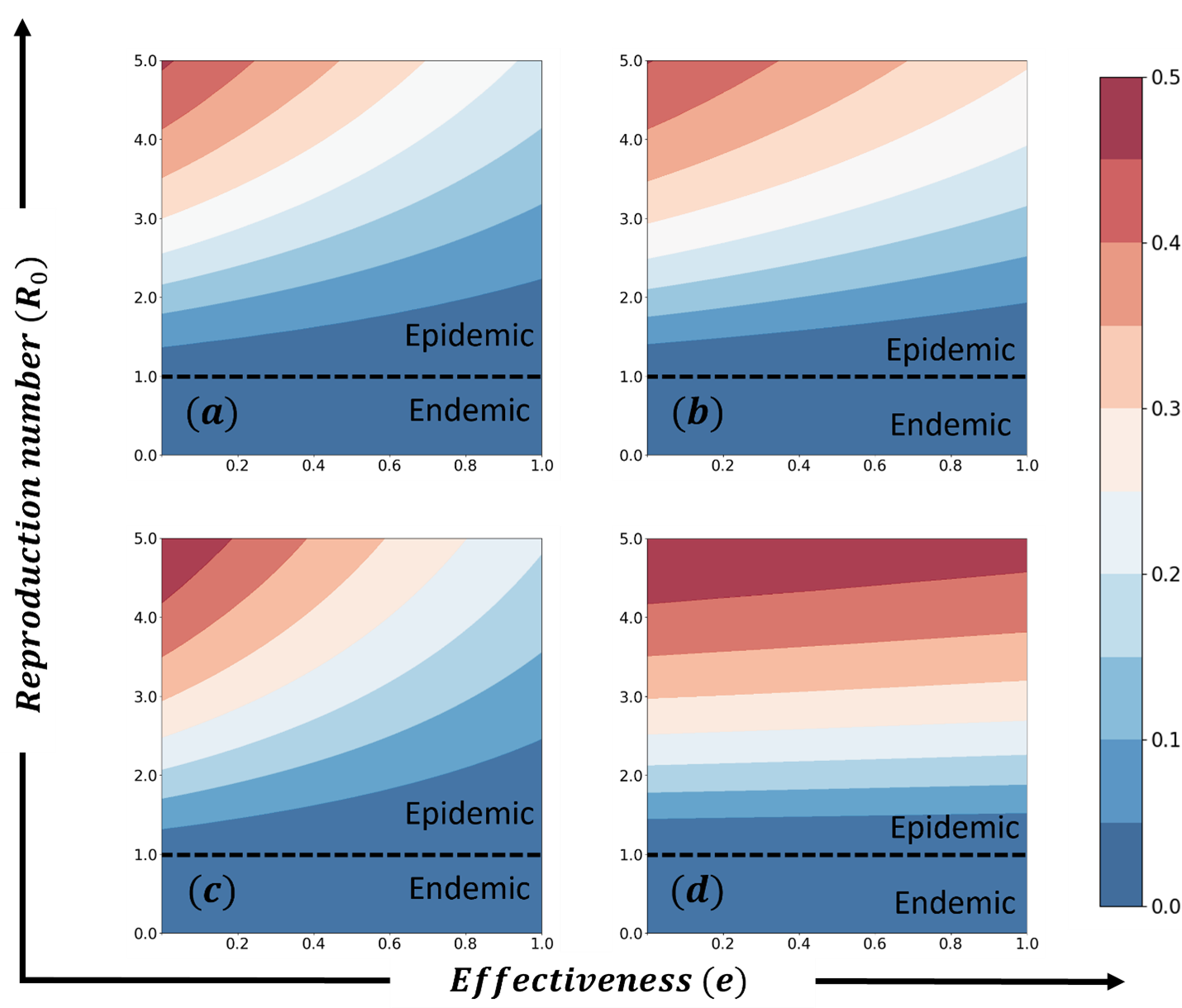}
	\caption{Heat map of the maximum infected population without control.}
	\label{fig:9}
\end{figure}

	\noindent In \textbf{Fig.\ref{fig:9}} the 'no control' baseline, evidently displays a threshold around $R_0=1$. Below this, it is Endemic (fewer infections, cool zones and blue colors); above, 'Epidemic' (more infections, hot zones and red/orange colors). This matches basic epidemiology: an infection requires $R_0>1$ to take off. The 'Effectiveness (e)' axis here probably represents intrinsic population susceptibility, rather than active control.
	
\begin{figure}[H]
		\centering
		\includegraphics[scale=1.15]{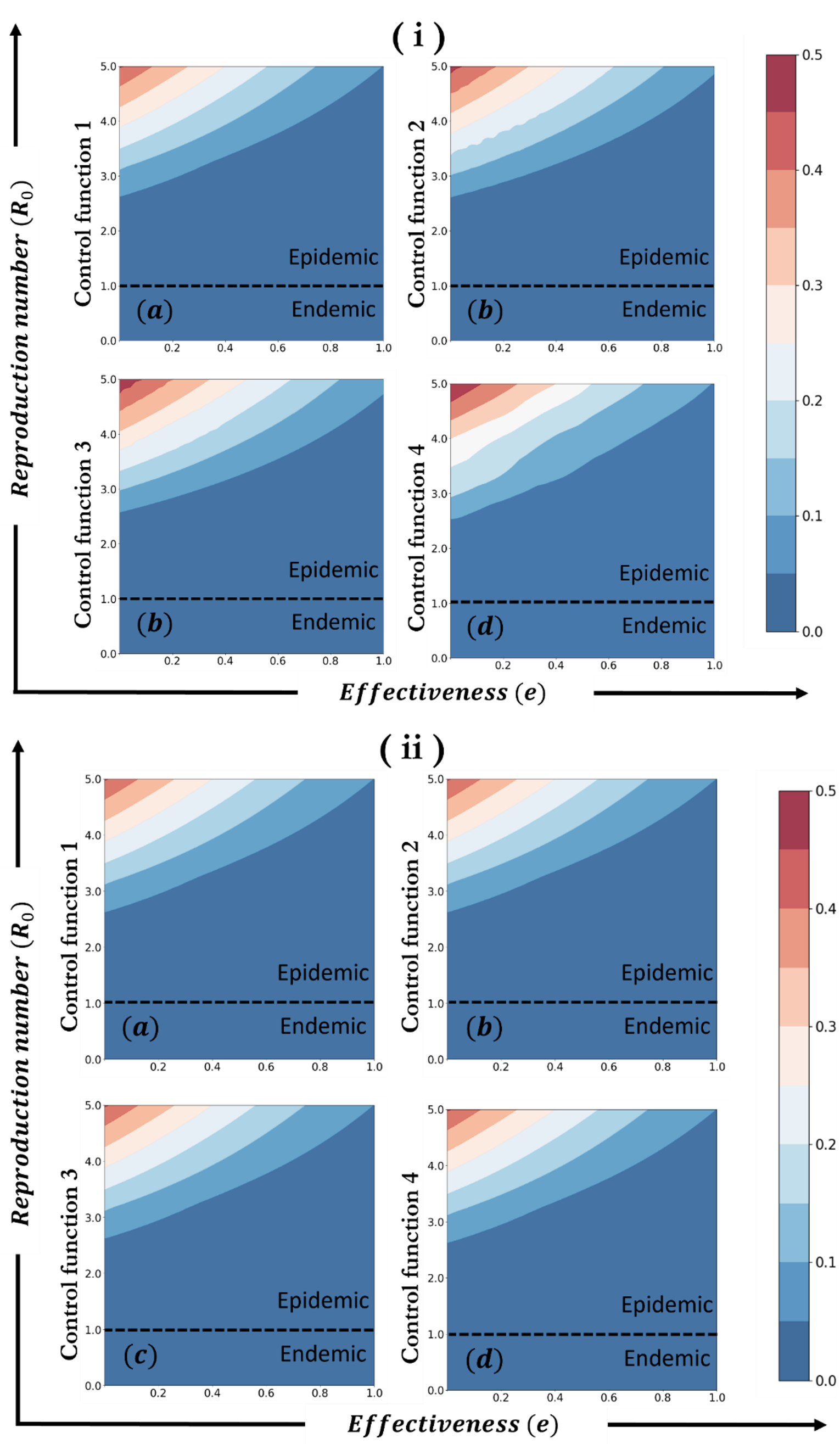}
		\caption{Heat map of the maximum infected population with control (i)$(\lambda=1$ and (ii) $\lambda=0.001)$.The initial fraction for all the four panels are (a) $S_0 = 0.55, V_0 = 0.31, I_0 = 0.14$, (b) $S_0 =0.55, V_0 = 0.40, I_0 = 0.05$, (c) $S_0 = 0.75, V_0 = 0.23, I_0 = 0.02$(d) $S_0 = 0.95, V_0 = 0.04, I_0 =0.01$}
		\label{fig:8}
	\end{figure}

	\noindent \textbf{Fig.\ref{fig:8}}, conversely, depicts scenarios with control, across two panels $(\lambda=1$ and $\lambda=0.001)$ and four distinct Control functions. These functions represent different intervention strategies (like quarantine or social distancing). Contrary to \textbf{Fig.\ref{fig:9}}  how increasing effectiveness directly reduces the infected population, shifting the system from 'Epidemic' to 'Endemic' even at higher $R_0$ values. This highlights the vital role of effective interventions for a few sets of initial conditions. 
	
\noindent The heatmaps also emphasize the importance of balancing vaccination coverage and effectiveness to effectively suppress outbreaks. For values of vaccination effectiveness greater than 0.6 and $R_0$ less than 1, the heatmaps consistently show blue zones where infections remain minimal. These regions indicate the onset of herd immunity, which is achieved when a critical fraction of the population becomes immune, reducing the effective reproduction number below 1. This critical threshold is often expressed as $H = 1 - \frac{1}{R_0}$ where H is the proportion of the population that must be immune to halt the disease's spread or simply the herd immunity \cite{Aguas2020}. Furthermore, targeting high-risk groups in scenarios with high $S_0$ and low $V_0$ can prevent initial surges in infections. The heatmaps highlight the critical role of vaccination in controlling epidemics and provide a quantitative framework for optimizing vaccination strategies to minimize infection peaks and effectively control the spread.

\begin{figure}[H]
		\centering
		\includegraphics[height=0.95\textheight]{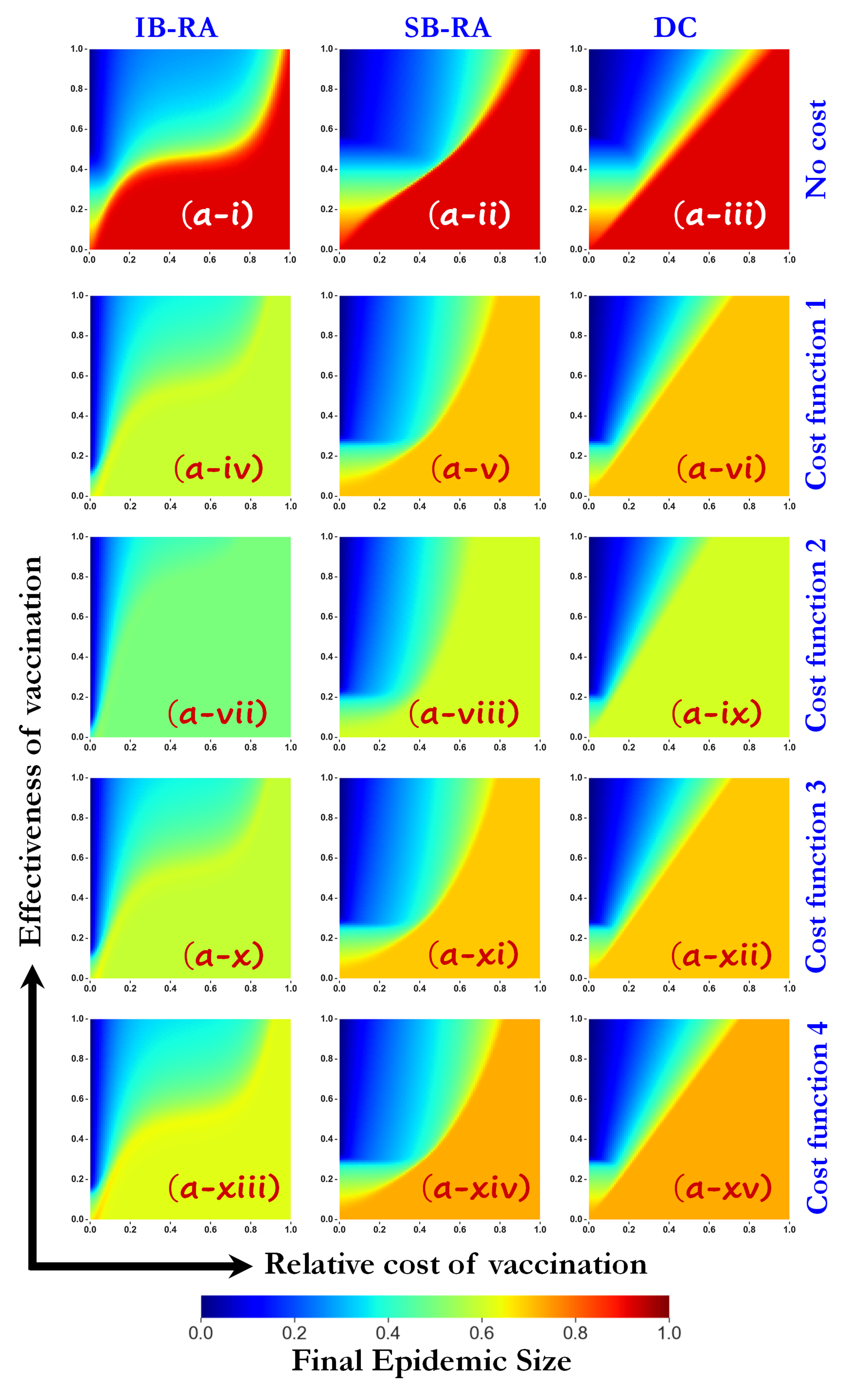}
		\caption{Final Epidemic Size (FES) for strategy-updating rule IB-RA (first column), SB-RA (second column), SB-RA (third column) for different cost functions cost function 1(first row),cost function 2(second row),cost function 3(third row),cost function 4(fourth row) }
		\label{fig:10}
	\end{figure}

\begin{figure}[H]
		\centering
		\includegraphics[height=0.95\textheight]{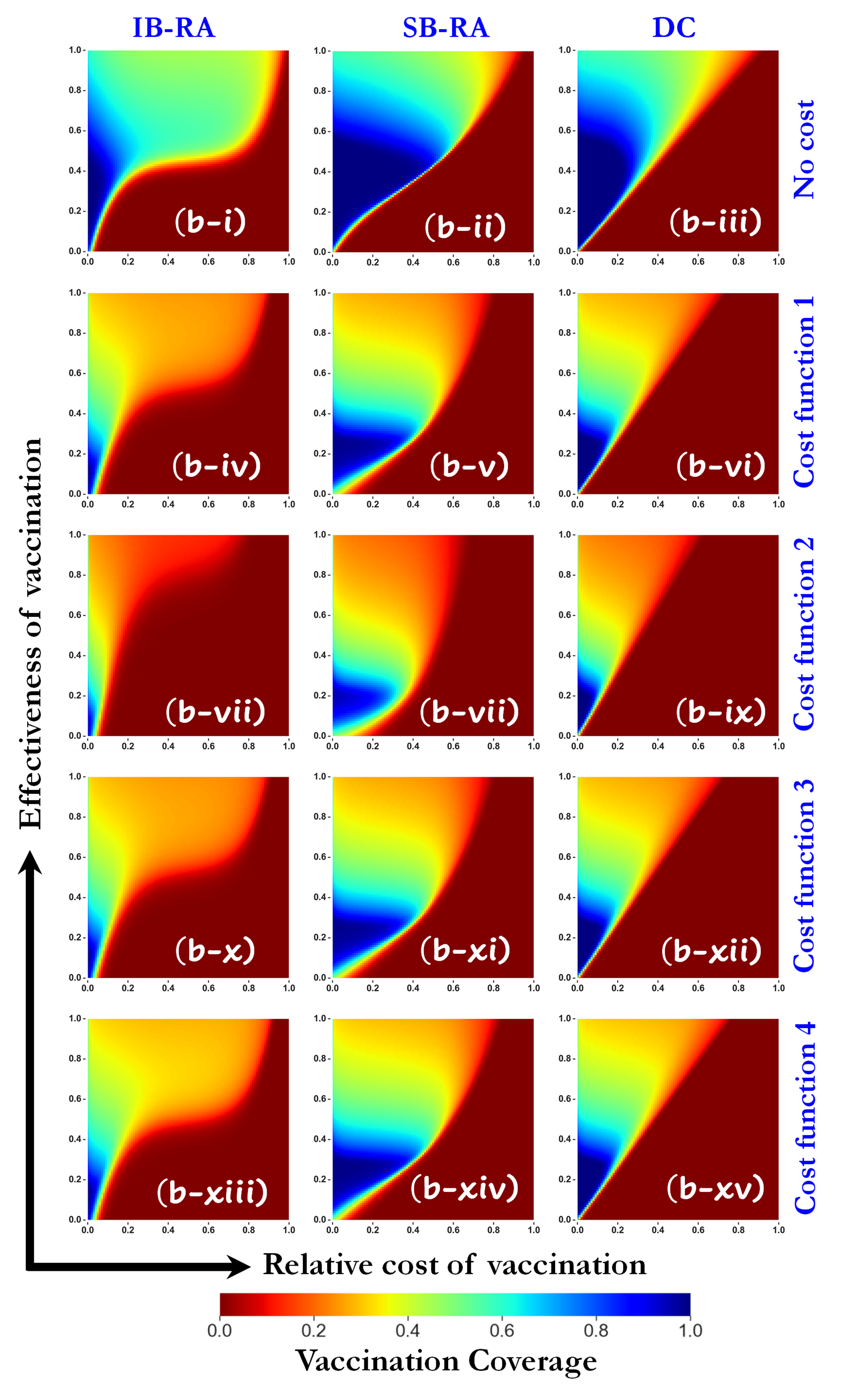}
		\caption{Vaccination Coverage (VC) for strategy-updating rule IB-RA (first column), SB-RA (second column), SB-RA (third column) for different cost functions cost function 1(first row),cost function 2(second row),cost function 3(third row),cost function 4(fourth row)}
		\label{fig:11}
	\end{figure}

\begin{figure}[H]
		\centering
		\includegraphics[height=0.95\textheight]{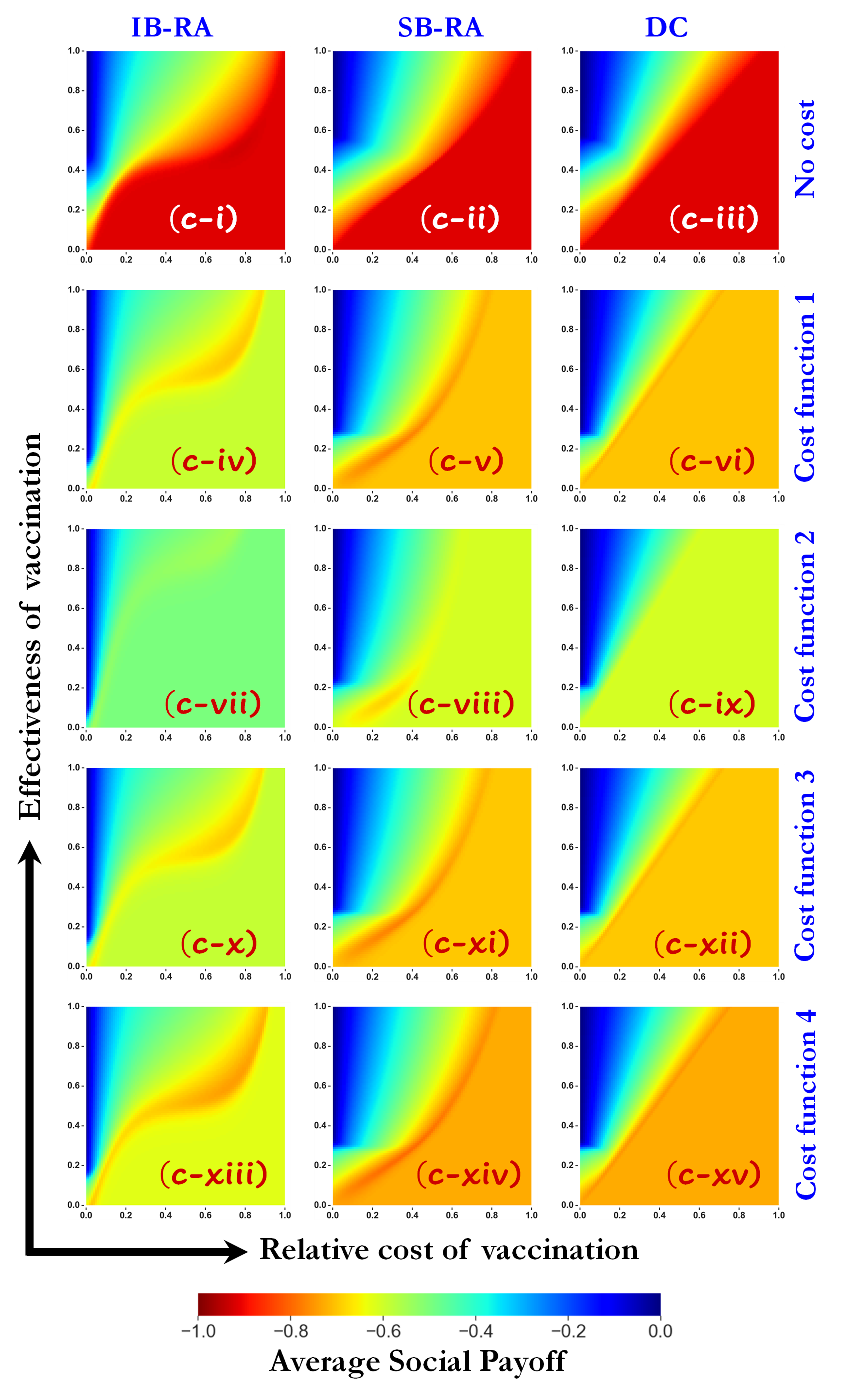}
		\caption{Average Social Payoff (ASP) for strategy-updating rule IB-RA (first column), SB-RA (second column), SB-RA (third column) for different cost functions cost function 1(first row),cost function 2(second row),cost function 3(third row),cost function 4(fourth row)}
		\label{fig:12}
	\end{figure}

  \noindent In Figures \ref{fig:10}--\ref{fig:12}, the dynamics of pandemic progression and the impact of optimal control strategies within the SIR/V framework under various strategy-updating rules within an epidemiological model are illustrated. These visuals are important to understand the impact of the optimal control on the epidemic. Figure \ref{fig:10} highlights scenarios both with and without control, showing the Final Epidemic Size (FES) under the strategy-updating rules: IB-RA (left column), SB-RA (middle column), and DC (right column). In all of these figures, the first row illustrates the pandemic without control, while the following four rows depict the impact of different control functions. The same organizational structure is maintained across all the figures. Similarly, Figure \ref{fig:11} illustrates Vaccination Coverage (VC), and Figure \ref{fig:12} presents the Average Social Payoff (ASP).\\
  
  \noindent In the heatmaps, the relative cost of vaccination is varied along the x-axis, while the effectiveness of the vaccine is varied along the y-axis. Each heatmap can be divided into two regions by the secondary diagonal: the upper triangle, where vaccine effectiveness exceeds the relative cost of vaccination, and the lower triangle, where the relative cost of vaccination exceeds vaccine effectiveness. In these heatmaps, preferred outcomes are represented by blue, indicating favorable scenarios, while worst-case situations are depicted in red, highlighting undesirable outcomes.\\

  \noindent When no control is applied, the heatmaps for Final Epidemic Size (FES) (Figure \ref{fig:10}), average social payoff (Figure \ref{fig:11}), and vaccination coverage (Figure \ref{fig:12}) exhibit a monotonic progression. The patterns appear smooth, showing a clear gradient between low and high values across vaccine effectiveness and the relative cost of vaccination. This indicates a steady relationship: as vaccine effectiveness increases, FES decreases, while social payoff and vaccination coverage increase reflecting natural behavioral responses in the absence of intervention.\\
  
  \noindent In contrast, when control is applied, the heatmaps exhibit non-monotonic behavior. The patterns become irregular, and the emergence of more complex regions reflects intricate interactions between vaccine effectiveness and cost. This shift arises from adaptive control measures, which are designed to optimize vaccination coverage while accounting for associated costs. Specifically, the application of control strategies dynamically adjusts vaccination uptake in response to changes in both effectiveness and cost. As a result, strategic behavior emerges, leading to fluctuations in vaccination coverage, social payoff, and epidemic size across various cost-effectiveness scenarios.\\
  
  \noindent This non-monotonic behavior can be attributed to several factors, as discussed in the literature \cite{Fenichel2011}. It highlights that adaptive human behavior in response to epidemic models can lead to complex, non-linear interactions between policy and individual decisions. Similarly, \cite{Dashtbali2020}  emphasizes that optimal control strategies in epidemic models, particularly those that adjust based on costs and effectiveness, often result in non-monotonic responses. \cite{Huang2022}  further elaborate that strategic behavior, such as choosing vaccination based on evolving conditions, can result in varying outcomes that are not predictable in a straightforward manner, especially when cost functions are considered.
  
  \noindent Applying the cost function $c(u)= -u \ln(1-u)$, the Final Epidemic Size (FES), average social payoff, and vaccination coverage exhibit marked improvements under the control strategies. The light red regions in Figure \ref{fig:10} specifically in panels (a-vii), (a-viii), and (a-ix) shift to light green under IB-RA and to light yellow under DC and SB-RA, indicating a reduced epidemic size and improved social outcomes. Additionally, the suppression  of blue areas in Figure \ref{fig:11} signifies decreased vaccination coverage. The shrinking dark blue regions underscore the effectiveness of strategic interventions in mitigating the spread of infection. These results highlight the pivotal role of cost-effective and adaptive control measures in reshaping epidemic trajectories and improving population health. 
  
  \noindent The red-shaded areas in all three figures FES, vaccination coverage, and average social payoff correspond to scenarios where the pandemic is ongoing and most of the population remains unvaccinated. More precisely, individuals are either not using partially effective vaccines or are unprotected against infection. As a result, widespread propagation of the disease occurs. Broadly speaking, these red regions emerge when vaccine effectiveness is very low or when vaccination requires high expenditure. This outcome is expected, as people tend to avoid vaccination when it is perceived as either unreliable or too costly.
  
  \noindent In scenarios where the relative cost of vaccination is low and vaccine effectiveness is high (depicted by the blue region in the upper triangular section), a larger proportion of the population is covered under the vaccination program. Consequently, the final epidemic size is significantly reduced, and the average social payoff approaches zero. This outcome is consistently represented by the blue regions in the upper triangles of all heatmaps for Final Epidemic Size (FES), Vaccination Coverage (VC), and Average Social Payoff (ASP).
  
  \noindent The boundaries between uniform and transitional regions in the heatmaps represent critical thresholds in vaccine effectiveness and cost—conditions that determine whether the epidemic remains uncontrolled or enters a managed state. These boundaries indicate a phase transition between the pandemic and controlled regimes. In Figure \ref{fig:10}, the lighter and darker red regions in panels (a-i), (a-ii), and (a-iii) indicate widespread infections and low vaccination uptake. These outcomes are driven by low vaccine effectiveness or high vaccination costs, reflecting public hesitance to adopt costly or unreliable vaccination measures. The boundary zones between uniform areas and the surrounding regions represent critical thresholds of vaccine effectiveness and cost, marking a phase transition from uncontrolled outbreaks to managed scenarios.

  \noindent Interestingly, within the controlled region, lower vaccine effectiveness can sometimes lead to increased vaccination coverage, particularly when the cost is low. While reduced cost encourages greater uptake, it is important to note that even widespread vaccination cannot eradicate the epidemic if vaccine effectiveness remains low. However, when vaccine efficacy is too low, it prevents complete eradication of the epidemic.
  
  \noindent Overall, These control strategies effectively mitigate the impact of the pandemic by reducing the final epidemic size and optimizing the average social payoff. The impact of different optimal control functions is clearly visible in Figures \ref{fig:10}, \ref{fig:11}, and \ref{fig:12}. In these figures, regions that appeared darker when no control was applied become lighter under control strategies, as expected—reflecting improved outcomes. As different types of control functions have been selected, their impacts vary depending on the strategy adaptation. Specifically, the control function performs more effectively under IB-RA when vaccine effectiveness is lower than the relative cost. In contrast, it is more effective under SB-RA and DC when vaccine effectiveness exceeds its relative cost.

\section{Conclusion}
\label{Conclusion}

\noindent In this study, we unified theoretical analysis with a thorough numerical investigation to address an adaptive optimal control problem for managing the early stage of an infectious disease outbreak. Assuming an unlimited, homogeneously mixed population, we constructed an analytical framework to model a vaccination strategy in an SIR/V compartmental structure. Three distinct strategy-updating mechanisms were considered, each executed sequentially. These mechanisms guided the application of optimal control functions designed to mitigate the spread of infection before it escalated. The primary focus of the study was on developing a pre-immunization strategy aimed at reducing the susceptible population through vaccination. This preemptive strategy altered the evolutionary dynamics of the outbreak when additional factors, such as partial vaccination coverage or defenses against infection, were introduced to halt the spread of the epidemic. We implemented and compared two scenarios: one with optimal control strategies applied to minimize infection, and one without any control measures. By conducting numerical experiments and creating schematic comparisons between the different strategies, we were able to derive adaptive optimal control functions and the corresponding evolutionary dynamics of the system. The results demonstrated that applying optimal control measures significantly reduced the infected population, underscoring the importance of well-planned vaccination strategies in curbing the spread of infectious diseases.

\section*{Funding}
This research is partially funded by the University Grants Commission
(UGC), Bangladesh, and the University of Dhaka, Bangladesh. This research did not receive any specific grant from funding agencies in the commercial or not-for-profit sectors.

\section*{Authors' contribution}
\label{Authors' contribution}

Nuruzzaman Rahat conceptualized the model, carried out the numerical simulations to validate the proposed model, visualized the model outcomes, made the original draft and critically revised the manuscript. Abid Hossain developed the model, carried out numerical simulations, analyzed the outcomes, made a contribution to the original draft. Muntasir Alam reviewed the manuscript, summarized the results, and supervised the project.

\section*{Acknowledgements}
The authors would like to thank the University Grants Commission (UGC), Bangladesh, and the Department of Applied Mathematics, University of Dhaka for their continued support and encouragement.

\section*{Declaration of competing interest}
The authors have no conflicts of interest to declare. We clarify that the submission is original work and is not under review at any other publication.

\section*{Data sharing}
There is no available data regarding this study. We do not analyze or generate any datasets, because our work proceeds within a theoretical and mathematical approach.

\section*{Ethical approval}
No consent is required to publish this manuscript. All authors have given approval for publication and agreed to be held accountable for the work performed herein.

\bibliographystyle{unsrt}

\bibliography{./mybib}



	\newpage

\end{document}